\documentclass[10pt,a4paper]{article}
\pdfoutput=1
\usepackage{jhepmod}

\usepackage{amsmath}
\usepackage{amssymb}
\usepackage{graphicx,color}

\usepackage{amsmath,amsthm}
\usepackage{txfonts}

\usepackage{bm}

\newtheorem{Theorem}{\underline{Theorem}}[section]
\newtheorem{Definition}{\underline{Definition}}[section]
\newtheorem{Lemma}[Theorem]{\underline{Lemma}}%[section]

\newtheorem{Proposition}[Theorem]{\underline{Proposition}}%[section]
\newcommand{\startproof}{\textbf{\underline{Proof}:\ \ }}
\newcommand{\finishproof}{\hfill $\Box$ \\}
\newcommand{\Z}{\mathbb{Z}}
\newcommand{\R}{\mathbb{R}}
\newcommand{\C}{\mathbb{C}}

\newcommand{\half}{\frac{1}{2}}
\newcommand{\Slc}{\mathrm{SL}(2,\mathbb{C})}
\newcommand{\slc}{\fs\fl_2\mathbb{C}}

\newcommand{\Su}{\mathrm{SU}(2)}

\def\be{\begin{eqnarray}}
\def\ee{\end{eqnarray}}

\newcommand{\cg}{\mathcal G}

\newcommand{\ck}{\mathcal K}

\newcommand{\cm}{\mathcal M}

\newcommand{\calr}{\mathcal R}

  \newcommand{\Fj}{\mathfrak{J}}
  
\newcommand{\fl}{\mathfrak{l}}

  \newcommand{\Fp}{\mathfrak{P}}

\newcommand{\fs}{\mathfrak{s}}

\renewcommand{\a}{\alpha}
\renewcommand{\b}{\beta}
\newcommand{\g}{\gamma}

\newcommand{\eps}{\varepsilon}

\newcommand{\sig}{\sigma}

\renewcommand{\l}{\lambda}
\renewcommand{\L }{\Lambda}
\renewcommand{\o}{\omega}
\renewcommand{\O}{\Omega}

\newcommand{\rmd}{\mathrm d}

\newcommand{\lt}{\left}
\newcommand{\rt}{\right}

\newcommand{\lag}{\left\langle}
\newcommand{\rag}{\right\rangle}

\newcommand{\tr}{\mathrm{tr}}
\newcommand{\bbc}{\mathbb{C}}

\newcommand{\id}{\mathrm{id}}

\newcommand{\bp}{\mathbf{P}}
\newcommand{\act}{\rhd}
\newcommand{\background}{(\mathring{j}_f,\mathring{g}_{ve},\mathring{z}_{vf})}

\title{Path Integral Representation of Lorentzian Spinfoam Model, Asymptotics, and Simplicial Geometries}

\author[]{Muxin Han,\ \ \ Thomas Krajewski}

\affiliation[]{Centre de Physique Th\'eorique%
\footnote{Unit\'e mixte de recherche (UMR 6207) du CNRS et des Universit\'es
de Provence (Aix-Marseille I), de la Meditarran\'ee (Aix-Marseille II) et du Sud (Toulon-Var); laboratoire affili\'e \`a la FRUMAM (FR 2291).}, CNRS-Luminy, Case 907, 13288 Marseille, France}

\emailAdd{Muxin.Han(AT)cpt.univ-mrs.fr} \emailAdd{Thomas.Krajewski@cpt.univ-mrs.fr} %\emailAdd{cuthor@another.univ.country}

\abstract{A new path integral representation of Lorentzian Engle-Pereira-Rovelli-Livine (EPRL) spinfoam model is derived by employing the theory of unitary representation of $\Slc$. The path integral representation is taken as a starting point of semiclassical analysis. The relation between the spinfoam model and classical simplicial geometry is studied via the large spin asymptotic expansion of the spinfoam amplitude with all spins uniformaly large. More precisely in the large spin regime, there is an equivalence between the spinfoam critical configuration (with certain nondegeneracy assumption) and a classical Lorentzian simplicial geometry. Such an equivalence relation allows us to classify the spinfoam critical configurations by their geometrical interpretations, via two types of solution-generating maps. The equivalence between spinfoam critical configuration and simplical geometry also allows us to define the notion of globally oriented and time-oriented spinfoam critical configuration. It is shown that only at the globally oriented and time-oriented spinfoam critical configuration, the leading order contribution of spinfoam large spin asymptotics gives precisely an exponential of Lorentzian Regge action of General Relativity. At all other (unphysical) critical configurations, spinfoam large spin asymptotics modifies the Regge action at the leading order approximation. }

\keywords{Covariant Loop Quantum Gravity, Lattice Models of Gravity, Models of Quantum Gravity}

\arxivnumber{}

\begin{document}

\maketitle%\vspace{-7mm}

\section{Introduction}

Loop Quantum Gravity (LQG) is an attempt to make a background independent, non-perturbative
quantization of 4-dimensional General Relativity (GR) -- for reviews, see \cite{book,rev,sfrevs}. It is
inspired by the classical formulation of GR as a dynamical theory of connections. The present article is concerning the covariant spinfoam approach of LQG \cite{sfrevs}. The current spinfoam models for quantum gravity are mostly inspired by the 4-dimensional Plebanski formulation of GR (or Plebanski-Holst formulation by including the Barbero-Immirzi parameter $\g$), which is a BF theory constrained by the condition that the $B$ field should be ``simple'' i.e. there is a tetrad field $e^I$ such that $B=*(e\wedge e)$. Currently one of the successful spinfoam models is the Engle-Pereira-Rovelli-Livine (EPRL) spinfoam model defined in \cite{EPRL,QSF}, whose (weak) implementation of simplicity constraint is understood in the sense of \cite{DingYou}.

The semiclassical behavior of spinfoam model is currently understood in terms of the \emph{large spin asymptotics} of the spinfoam amplitude. We consider a spinfoam model as a state-sum
\be
A(\ck)=\sum_{j_f}\mu(j_f)A_{j_f}(\ck)
\ee
where $\mu(j_f)$ is a weight. We investigate the asymptotic behavior of the (partial-)amplitude $A_{j_f}$ as all the spins $j_f$ are taken to be large uniformly. The area spectrum in LQG is given approximately by $A_f=\g j_f\ell_p^2$, so the semiclassical limit of spinfoam models is argued to be achieved by taking $\ell_p^2\to0$ while keeping the area $A_f$ fixed, which results in $j_f\to\infty$ uniformly as $\g$ is a fixed Barbero-Immirzi parameter. There is another argument relating the large-j asymptotics of the spinfoam vertex amplitude to the semiclassical limit, by imposing the semiclassical boundary state to the vertex amplitude \cite{holomorphic}. Mathematically the asymptotic problem is posed by making a uniform scaling for the spins $j_f\mapsto\l j_f$, and studying the asymptotic behavior of the amplitude $A_{\l j_f}(\ck)$ as $\l\to\infty$. 

In the above sense of large-j asymptotics, the EPRL vertex amplitude is shown to reproduce the classical discrete GR in the large-j asymptotics \cite{semiclassical}. More precisely it is shown that the asymptotics of the EPRL vertex amplitude is mainly a Cosine of the Regge action in a 4-simplex if the boundary data admits a nondegenerate 4-simplex geometry, and the asymptotics is non-oscillatory if the boundary data doesn't admit a nondegenerate 4-simplex geometry. There are many investigations for the large-j asymptotics of various spinfoam models. For example, the asymptotics of the Barrett-Crane vertex amplitude (10j-symbol) is studied in \cite{10j}, which shows that the spinfoam configurations of degenerate geometry in Barrett-Crane model were non-oscillatory, but dominant. The large-j asymptotics of the FK model is studied in \cite{CF}, concerning the nondegenerate Riemanian geometry, in the case of a simplicial complex without boundary. A general investigation is carried out in \cite{HZ} for the EPRL model of both Euclidean and Lorentzian signature on a simplicial complex either with or without boundary, including the analysis of spinfoam configurations of both nondegenerate and degenerate geometries. There are also studies of the Euclidean/Lorentzian spinfoam amplitude by including an additional scaling of Barbero-Immirzi parameter \cite{claudio}.

In the present work, we develop a new path integral representation of the Lorentzian EPRL spinfoam amplitude on a simplicial manifold. The path integral representation is derived from top to down from the group-representation-theoretic definition of the model in \cite{Carlo}, it is more elegant and economic than the one employed in \cite{HZ} because it has less integration variables. The derivation toward a path integral representation uses the theory of unitary representations (principle series) of $\Slc$, which is summarized in Appendix \ref{Slc}. The new path integral representation gives a new spinfoam action, which controls the large spin asymptotics of the spinfoam amplitude via the stationary phase approximation.

Here we still focus on the discussion of spinfoam partial amplitude. When the sum over spin is taking into account, the semiclassical behavior of the spinfoam model is investigated in the companion papers \cite{statesum,han}.

In the present paper we develop a systematic analysis of the spinfoam large spin asymptotics. We make the discussion pedagogical and self-contained in this paper. Here we clarify the relation between the spinfoam model and classical simplicial geometry via the large spin asymptotic expansion. More precisely, in the large spin regime, there is an equivalence between the spinfoam critical configuration (with certain nondegeneracy assumption) and a classical Lorentzian simplicial geometry (discussed in Section \ref{GeoEq}). Such an equivalence relation allows us to classify the spinfoam critical configurations by their geometrical interpretations, via two types of solution-generating maps (in Section \ref{SG}). The equivalence between spinfoam critical configuration and simplical geometry also allows us to define the notion of globally oriented and time-oriented spinfoam critical configuration (in Section \ref{timeorientation}). It is shown (in Section \ref{CriticalAction}) that only at a globally oriented and time-oriented spinfoam critical configuration, the leading order contribution of spinfoam large spin asymptotics gives precisely an exponential of Lorentzian Regge action of General Relativity. At all other critical configurations, spinfoam large spin asymptotics modifies the Regge action at the leading order approximation.

 \section{Lorentzian EPRL Spinfoam Model and Path integral Representation}

We start with a simplicial complex $\ck$ whose dual skeleton is a 2-complex ${\cal C}$ made of vertices $V$, edges $E$ and faces $F$. The 2-complex may have a boundary $\partial{\cal C}$ which is a graph $\Gamma$ whose vertices $N$ and edges $L$ we call nodes and links. Edges and vertices of ${\cal C}$ not in $\partial{\cal C}$ are call internal. The incidence matrices between faces and edges $\epsilon_{f,e}$ and between edges and vertices $\epsilon_{e,v}$ are defined to be $\pm1$ depending on whether the corresponding lower  dimensional cell of ${\cal C}$ ($e$ or $v$) is on the boundary of the higher dimensional one ($f$ or $e$) with  an orientation that agrees ($+1$) or disagrees ($-1$). If the there is no incidence relation between the cells, the matrix element of $\epsilon$ vanishes.

Following the definition in \cite{Carlo}, we associate $h_{f,e}\in \Su$ to the strands (edges that are split in as many faces they meet) and $g_{e,v}\in \Slc$ to the half edges. Each face is associated with a spin $j_{f}\neq 0$. Then the Lorentzian EPRL spinfoam amplitude \cite{EPRL} of the simplicial complex $\ck$ reads (we assume momentarily that $\partial{\ck}=\emptyset$, but the general case can be treated, see \cite{Carlo}),
\begin{multline}
{A}(\ck)=\int_{\Slc} \prod \rmd g_{e,v}\\\prod_{f}\Bigg\{ \sum_{j_{f}\neq 0}\underbrace{d_{j_{f}}\int_{\Su}\prod_{e\in\partial f} \rmd h_{f,e}
\mbox{Tr}_{{\cal H}_{j_{f}\!,\gamma j_{f}}}
\Big[\mathop{\prod}\limits^{\longrightarrow}_{e\in\partial f}\big(g_{e,s(e)}\,h_{f,e}\,g_{e,t(e)}^{-1}\big)^{\epsilon_{f,e}}\Big]
\mathop{\prod}\limits_{e\in\partial f}d_{j_f}\mbox{Tr}_{{\cal H}_{j_{f}}}\Big[h_{f,e}\Big]}_{\mathrm{face\, amplitude}\,{\cal A}_{f}}\Bigg\},
\end{multline}
with $s(e)$ and $t(e)$ the source and the target of $e$. The arrows on the products means that they have to be ordered following the orientation of the face. $d_j=2j+1$ is the dimension of SU(2) spin-$j$ representation. The spinfoam model $A(\ck)$ written above is invariant under flipping of the edge and face orientations \cite{eforientation}.

Let us focus on a single face and drop the face index $f$ temporarily to alleviate the notations. For any vertex $v$, let $e'(v)$ and $e''(v)$ be the edge that enters (resp. leaves) $v$ following the orientation of the the face. Since these edges depend on the face, one should denote them by $e'_{f}(v)$ and $e''_{f}(v)$. 
Then, the contribution of the face $f$ to the amplitude reads (to prove this result, simply distinguish the four cases depending on the orientations of the edges incident to $v$ in $f$),
\be
{\cal A}_{f}=\sum_{j\neq 0}d_{j}\int_{SU(2)}\prod_{e\in\partial f} dh_{e}
\mbox{Tr}_{{\cal H}_{j\!,\gamma j}}
\Big[\mathop{\prod}\limits_{v\in\partial f}^{\longrightarrow}g^{-1}_{e'(v),v}g_{e''(v),v}h_{e''(v)}\Big]
\mathop{\prod}\limits_{e\in\partial f}d_{j}\mbox{Tr}_{{\cal H}_{j}}\big[\,\overline{h}_{e}\big].
\ee
We have also used the reality of the $\Su$ characters $\mbox{Tr}_{{\cal H}_{j}}[h]=\mbox{Tr}_{{\cal H}_{j}}[\,\overline{h}\,]$ to introduce complex conjugates in the second factor and to change $h^{-1}_{e}$ into $h_{e}$ whenever the orientation of $f$ and $e$ disagree.

Because any $h_{e}$ appears only twice, the integration over $h_{e}$ can be performed using Schur's orthogonality relation \eqref{Schur},
\begin{equation}
{\cal A}_{f}=\sum_{j\neq 0}d_{j}
\mbox{Tr}_{{\cal H}_{j,\g j}}
\Big[\mathop{\prod}\limits_{v\in\partial f}^{\longrightarrow}P_{j}\,g^{-1}_{e'(v),v}g^{\phantom{e''(v),v}}_{e''(v),v}\,P_{j}\Big],
\end{equation}
with $P_{j}$ the orthogonal projector onto the spin $j$ representation ${\cal H}_{j}$ of $SU(2)$ contained in ${\cal H}_{j,\gamma j}$. 

Using the canonical basis \eqref{basis} of ${\cal H}_{j\!,\gamma(j+1)}$, the projector can be expressed as
\begin{equation}
P_{j}=\sum_{-j\leq m\leq j}|j.m\rangle_{{\cal H}_{j\!,\gamma j}}{}_{{\cal H}_{j\!,\gamma j}}\langle j,m|.
\end{equation}
Then, at each vertex $v$ on the boundary of $f$ we have a scalar product
\begin{equation}
{}_{{\cal H}_{j,\gamma j}}\langle j,m_{e'}|g^{-1}_{e',v}g^{\phantom{e'',v}}_{e'',v}|j,m_{e''}\rangle_{{\cal H}_{j,\gamma j}},
\end{equation}
Accordingly, the basic building blocks of the amplitude are the scalar products
\begin{equation}
{}_{{\cal H}_{j,\gamma j}}\langle j,m|g^{-1}g'|j,m'\rangle_{{\cal H}_{j,\gamma j}},
\end{equation}
Using the unitarity of the representation, this is nothing but the $L^{2}(SU(2))$ scalar product of $g|j,m\rangle_{{\cal H}_{j,\gamma j}}$ with $g'|j,m'\rangle_{{\cal H}_{j,\gamma j}}$, so that it may be expressed in terms of Wigner matrices as follows,
\be
{}_{{\cal H}_{j,\gamma j}}\langle j,m|g^{-1}g'|j,m'\rangle_{{\cal H}_{j,\gamma j}}=
d_{j}\,\int_{SU(2)}\!dh\,\,\, [\lambda_{g}(h)]^{-2\mathrm{i}\gamma j-2}[\lambda_{g'}(h)]^{2\mathrm{i}\gamma j-2}
\overline{D_{j,m}^{j}\big[h_{g}(h)\big]}D_{j,m'}^{j}\big[h_{g'}(h)\big].
\ee
Because of the unitarity of the Wigner matrices, this also reads
\begin{equation}
d_{j}\,\int_{SU(2)}\!dh\,\,\, [\lambda_{g}(h)]^{-2\mathrm{i}\gamma j-2}[\lambda_{g'}(h)]^{2\mathrm{i}\gamma j-2}
\langle j,m|h^{\dagger}_{g}(h)|j,j\rangle\langle j,j| h_{g'}(h)|j,m'\rangle,
\end{equation}
with all matrix elements taken in the spin $j$ representation of $SU(2)$.

Turning back to the face amplitude, we get rid of the summation over magnetic numbers using  the resolution of the identity, so that the face amplitude reads
\be
{\cal A}_{f}&=&\sum_{j}d_{j}^{|V(f)|+1}\int_{SU(2)}\prod_{v\in \partial f}dh_{v}\,
\prod_{e\in\partial f}\Big\{[\lambda_{g_{e'',v}}(h_{v})]^{-2\mathrm{i}\gamma j-2}
[\lambda_{g_{e',v}}(h_{v})]^{2\mathrm{i}\gamma j-2}\Big\}\\
&&\times\mbox{Tr}_{{\cal H}_{j}}
\Big[\mathop{\prod}\limits_{v\in\partial f}^{\longrightarrow}\,h^{\dagger}_{g_{e',v}}(h_{v})|j,j\rangle\langle j,j|h_{g_{e'',v}}(h_{v})\Big]
\ee
where $|V(f)|$ is the number of vertices along $\partial f$. To express the face amplitude in terms of coherent states, we have to compute the matrix element
\begin{align}
\langle j,j|h_{g'}(h')h^{\dagger}_{g}(h)|j,j\rangle=
\langle \textstyle{\frac{1}{2},\frac{1}{2}}|h_{g'}(h')h^{\dagger}_{g}(h)|\textstyle{\frac{1}{2},\frac{1}{2}}\rangle^{2j}
\end{align}
Since $h_{g}(h)$ is defined by the decomposition $hg=k_{g}(h)h_{g}(h)$, we have
\begin{equation}
h_{g}(h)=k_{g}^{-1}(h)hg=k_{g}^{\dagger}(h)h(g^{-1})^{\dagger}
\end{equation}
as well as the conjugate
\begin{equation}
h^{\dagger}_{g}(h)=g^{\dagger}h^{\dagger}(k_{g}^{-1})^{\dagger}(h)=(g^{-1})h^{\dagger}k_{g}(h).
\end{equation}
Because
\begin{equation}
k_{g}(h)=\begin{pmatrix}\lambda_{g}^{-1}(h)&\mu_{g}(h)\cr 0&\lambda_{g}(h)\end{pmatrix},
\end{equation}
$|\frac{1}{2},\frac{1}{2}\rangle$ is an eigenvector of $k_{g}(h)$ with eigenvalue $\lambda^{-1}_{g}(h)$. Thus the matrix element reads
\begin{align}
\langle \textstyle{\frac{1}{2},\frac{1}{2}}|h_{g'}(h')h^{\dagger}_{g}(h)|\textstyle{\frac{1}{2},\frac{1}{2}}\rangle=
\lambda_{g}^{-1}(h)\lambda_{g'}^{-1}(h')\langle \textstyle{\frac{1}{2},\frac{1}{2}}|h' ((g'^{-1})^{\dagger} g^{-1}h^{\dagger}|\textstyle{\frac{1}{2},\frac{1}{2}}\rangle
\end{align}
Thus, the face amplitude reads
\be
{\cal A}_{f}=\sum_{j}d_{j}^{|V(f)|+1}\int_{SU(2)}\prod_{v\in \partial f}\rmd h_{v}\,
\mbox{Tr}_{{\cal H}_{j}}
\Bigg[\mathop{\prod}\limits_{v\in\partial f}^{\longrightarrow}\,
\frac{g^{-1}_{e',v}h_{v}^{\dagger}|j,j\rangle\langle j,j|h_{v}(g_{e'',v}^{-1})^{\dagger}}
{[\lambda_{g_{e',v}}(h_{v})]^{2 j(\mathrm{i}\gamma+1)+2}
[\lambda_{g_{e'',v}}(h_{v})]^{2 j(-\mathrm{i}\gamma+1)+2}}\Bigg]
\ee
Using \eqref{lambda}, we express the face amplitude in terms of the coherent state $|z_{v}\rangle=h_{v}^{\dagger}|\frac{1}{2},\frac{1}{2}\rangle$,
\be
{\cal A}_{f}=\sum_{j}d_{j}^{|V(f)|+1}\int_{\mathbb{CP}^1}\prod_{v\in \partial f}\rmd z_{v}\,
\frac{\Big\{\mbox{Tr}
\Big[\mathop{\prod}\limits_{v\in\partial f}^{\longrightarrow}\,
g^{-1}_{e',v}|z_{v}\rangle\langle z_{v}|(g_{e'',v}^{-1})^{\dagger}\Big]\Big\}^{2j}}
{\mathop{\prod}\limits_{v\in\partial f}
\langle z_{v}|(g_{e',v}^{-1})^{\dagger}g_{e',v}^{-1}|z_{v}\rangle^{ j(\mathrm{i}\gamma+1)+1}
\langle z_{v}|(g_{e'',v}^{-1})^{\dagger}g_{e'',v}^{-1}|z_{v}\rangle^{ j(-\mathrm{i}\gamma+1)+1}}
\ee
with the trace taken in the fundamental representation. Let us note that the coherent states always appear in pairs $|z_{v}\rangle\langle z_{v}|$ so that the phase ambiguity is irrelevant and we restrict the integration form $\Su$ to $S^{2}=\mathbb{CP}^1=(\C^2\setminus\{0\})/C^*$. $\rmd {z}=\frac{i}{2}\left( z_{0}dz_{1}-z_{1}dz_{0}\right) \wedge \left(
\bar{z}_{0}d\bar{z}_{1}-\bar{z}_{1}d\bar{z}_{0}\right)$ is a homogeneous measure of degree 4 on $\mathbb{C}^2$. The integration is essentially on $\mathbb{CP}^1$ since the integrand is a homogeneous function of $z_{vf}$ of degree $-4$.

Now we put back the label $f$ to the variables $j\mapsto j_f,z_v\mapsto z_{vf}$ and write $g_{e,v}=(g_{ve}^\dagger)^{-1}$. The $\Slc$ Haar measure $\rmd g_{e,v}=\rmd g_{ve}$\footnote{The $\Slc$ Haar measure $\rmd g$ can be written explicitly by
\be
\rmd g=\frac{\rmd\b\rmd{\b}^*\rmd\g\rmd{\g}^*\rmd\delta\rmd{\delta}^*}{|\delta|^2}\ \ \ \ \ g=\left(
                                                                                       \begin{array}{cc}
                                                                                        {\a} & \b \\
                                                                                        \g & \delta \\
                                                                                       \end{array}
                                                                                     \right)
\ee
which is manifestly invariant under $g\mapsto g^\dagger$.}. We define a new spinor variable $Z_{vef}$ and a degree 0 measure by
\begin{eqnarray}
Z_{vef} := g_{ve}^{\dag }z_{vf} \ \ \ \text{and}\ \ \ \ 
\Omega _{vf} := \frac{\rmd{z_{vf}}}{\left\langle
Z_{vef},Z_{vef}\right\rangle \left\langle Z_{ve^{\prime }f},Z_{ve^{\prime
}f}\right\rangle }
\end{eqnarray}%
Then the EPRL spinfoam amplitude $A(\ck)$ can be written as
\be
A(\ck)&=&\sum_{j_{f}}d_{j_f}^{|V(f)|+1} \int_{\Slc}\prod_{\left( v,e\right)}
\rmd g_{ve} \int_{\mathbb{CP}^{1}}\prod_{v\in \partial f} \rmd{\O_{vf}}\prod_{(e,f)}\frac{\lag Z_{vef},Z_{v'ef}\rag^{2j_f}}{ \lag Z_{v'ef},Z_{v'ef}\rag^{i\g j_f+j_f} \lag Z_{vef},Z_{vef}\rag^{-i\g j_f+j_f}}
\ee
where one may view that $v=s(e)$ is the source and $v'=t(e)$ is the target. The integrand can be written into an exponential form $e^S$ with an action $S$ written as
\be
S[j_f,g_{ve},z_{vf}]%&=&\sum_{(e,f)}j_f\ln\frac{\lag Z_{vef},Z_{v'ef}\rag^{2}}{ \lag Z_{v'ef},Z_{v'ef}\rag^{i\g +1} \lag Z_{vef},Z_{vef}\rag^{-i\g+1}}\nonumber\\
&=&\sum_{(e,f)}2j_f\lt[\ln\frac{\lag Z_{vef},Z_{v'ef}\rag}{ \lag Z_{v'ef},Z_{v'ef}\rag^{1/2} \lag Z_{vef},Z_{vef}\rag^{1/2}}+i\g \ln \frac{\lag Z_{vef},Z_{vef}\rag^{1/2}}{\lag Z_{v'ef},Z_{v'ef}\rag^{1/2}}\rt].\label{SSS}
\ee
The spinfoam action $S$ has the following continuous gauge degree of freedom:
\begin{itemize}
\item Rescaling of each $z_{vf}$\footnote{The measure $\rmd z_{vf}$ is scaling invariant.}:
\be
z_{vf}\mapsto \l z_{vf},\ \ \ \l\in\C.
\ee

\item $\Slc$ gauge transformation at each vertex $v$:
\be
g_{ve}\mapsto x^{-1}_vg_{ve},\ \ \ z_{vf}\mapsto x^\dagger_vz_{vf},\ \ \ x_v\in\Slc.
\ee

\item SU(2) gauge transformation on each edge $e$:
\be
g_{ve}\mapsto g_{ve}h_e^{-1},\ \ \ h_e\in\Su.
\ee

\end{itemize}
The spinfoam action $S$ has the following discrete gauge symmetry:

\begin{itemize}

\item Flipping the sign of the group variable $g_{ve}\mapsto -g_{ve}$. Thus the space of group variable is essentially the restricted Lorentz group $\mathrm{SO}^+(1,3)$ rather than its double-cover $\Slc$.

%\item $J$-parity as a $\Z_2$ symmetry acting simultaneously on $g_{ve}$ and $z_{vf}$:
%\be
%g_{ve}\mapsto J g_{ve} J^{-1}=(g_{ve}^\dagger)^{-1}, \ \ \ z_{vf}\mapsto Jz_{vf}
%\ee
%where $J$ is an anti-unitary map $J(z_0,z_1)^t:=(-\bar{z}_1,\bar{z}_0)^t$

\end{itemize}

%It is shown in the following that the $J$-parity symmetry is spontaneously broken in the regime that the spins $j_f$ are uniformly large.

The above derivation toward the spinfoam action $S$ gives a new path integral representation of the Lorentzian EPRL spinfoam model on a general simplicial complex (or even on an arbitrary cellular decomposition of the spacetime manifold with the generalization \cite{KKL,DingYou}) instead of the one presented in \cite{semiclassical,HZ}. The spinfoam action derived here has obvious advantage that it depends on less variables (only $j_f, g_{ve},z_{vf}$) than the one (depending on $j_f,g_{ve},z_{vf},\xi_{ef}$) in \cite{semiclassical,HZ}. %In the large spin asymptotic analysis of the spinfoam amplitude in \cite{HZ}, $\xi_{ef}$ are the degrees of freedom frozen by $z_{vf}$ from the critical equations. Thus the new spinfoam action $S$ is expected to remove some certain redundancies in the spinfoam action employed in \cite{HZ}. We come back to this point in Section \ref{redundant}.

For the convenience of the discussion, we define the notion of the partial amplitude $A_{j_f}(\ck)$ by collecting all the integrations
\be
A_{j_f}(\ck)&:=&\int
\rmd g_{ve} \int\rmd \O_{vf} e^{S[j_f,g_{ve},z_{vf}]}  \label{Aj}
\ee
So that the spinfoam state-sum is given by a sum of partial amplitude
\be
A(\ck)=\sum_{j_f}d^{|V(f)|+1}_{j_f}A_{j_f}(\ck).
\ee

\section{Large Spin Asymptotics}

In this paper, we consider the asymptotic behavior of the \emph{partial amplitude} $A_{j_f}(\ck)$ in the regime where all the spins are uniformly large. This asymptotic behavior can be studied by making a uniformly rescaling $j_f\mapsto \l j_f$ and assuming $\l\gg 1$. The asymptotic analysis may be viewed as the first step toward developing a perturbation theory of spinfoam model with respect to an arbitrary background configuration $({j}^0_f,{g}^0_{ve},z^0_{vf})$, where the background spins $j^0_f$ are uniformly large. The asymptotic analysis should show us which background configuration are preferred with relatively small fluctuations counted by $1/\l$. 

As a notation we write:
\be
{j}^0_f=\l \mathring{j}_f,\ \ \ {g}^0_{ve}\equiv \mathring{g}_{ve},\ \ \ z^0_{vf}\equiv\mathring{z}_{vf}
\ee
where $\l$ is a large parameter and $\mathring{j}_f\sim o(1)$. We sometimes refer $\l$ as the background spin in the following context, and we often refer $\background$ as the background configuration.

Because of the path integral representation of the spinfoam model, the asymptotic analysis of the partial amplitude $A_{\l j_f}(\ck)$ as $\l$ large is guided by the following general result (Theorem 7.7.5 and 7.7.1 in \cite{stationaryphase}):

\begin{Theorem}\label{asymptotics}
(A) Let $K$ be a compact subset in $\R^n$, $X$ an open neighborhood of $K$, and $k$ a positive integer. If (1) the complex functions $u\in C^{2k}_0(K)$, $S\in C^{3k+1}(X)$ and $\Re(S)\leq 0$ in $X$; (2) there is a unique point $x_0\in K$ satisfying $\Re(S)(x_0)=0$, $S'(x_0)=0$, and $\det S''(x_0)\neq 0$. $S'\neq0$ in $K\setminus \{x_0\}$, then we have the following estimation:
\be
\lt|\int_K u(x)e^{\l S(x)}\rmd x-e^{\l S(x_0)}\lt(\frac{2\pi}{\l}\rt)^{\frac{n}{2}}\frac{e^{\mathrm{Ind}(S'')(x_0)}}{\sqrt{\det(S'')(x_0)}}\sum_{s=0}^{k-1}\lt(\frac{1}{\l}\rt)^s L_s u(x_0)\rt|\leq C\lt(\frac{1}{\l}\rt)^{k+\frac{n}{2}}\sum_{|\a|\leq 2k}\sup\lt|D^\a u\rt|
\ee
Here the constant $C$ is bounded when $f$ stays in a bounded set in $C^{3k+1}(X)$. We have used the standard multi-index notation $\a =\lag\a_1,\cdots,\a_n\rag$ and
\be
D^\a=\frac{\partial^{|\a|}}{\partial x_1^{\a_1}\cdots\partial x_n^{\a_n}},\ \ \ \ \text{where}\ \ \ \ |\a|=\sum_{i=1}^n\a_i
\ee
$L_s u(x_0)$ denotes the following operation on $u$:
\be
L_s u(x_0)=i^{-s}\sum_{l-m=s}\sum_{2l\geq 3m}\frac{2^{-l}}{l!m!}\lt[\sum_{a,b=1}^nH^{-1}_{ab}(x_0)\frac{\partial^2}{\partial x_a\partial x_b}\rt]^l\lt(g_{x_0}^m u\rt)(x_0)\label{Lsu}
\ee
where $H(x)=S''(x)$ denotes the Hessian matrix and the function $g_{x_0}(x)$ is given by
\be
g_{x_0}(x)=S(x)-S(x_0)-\frac{1}{2}H^{ab}(x_0)(x-x_0)_a(x-x_0)_b
\ee
such that $g_{x_0}(x_0)=g_{x_0}'(x_0)=g_{x_0}''(x_0)=0$.

(B) Let $K$ be a compact subset in $\R^n$, $X$ an open neighborhood of $K$, $u\in C^{2k}_0(K)$ a compact support complex function, assuming $\Re(S)(x)\leq0$ in $X$ but the equations $\Re(S)(x)=0$, $S'(x)=0$ has no solutions in $K$, then there exists a constant $C$ such that 
\be
\lt|\int_K u(x)e^{\l S(x)}\rmd x\rt|\leq \lt(\frac{1}{\l}\rt)^kC\sum_{|\a|\leq k}\sup_K\frac{\lt|D^\a u\rt|}{\lt(|S'|^2+\Re(S)\rt)^{k-|\a|/2}}
\ee
for all $k\in\Z_+$.

\end{Theorem}

In Eq.\eqref{Lsu}, the expression of $L_s u(x_0)$ only sums a finite number of terms for all $s$. For each $s$, $L_s$ is a differential operator of order $2s$ acting on $u(x)$. %For example we list the possible types of terms in the sums corresponding to $s=1$ and $s=2$
%\begin{itemize}

%\item In the case $s=1$, the possible $(m,l)$ are $(m,l)=(0,1),(1,2),(2,3)$ to satisfy $2l\geq 3m$. The corresponding terms are of the types
%\be
%(m,l)=(0,1):&& \partial^2u(x_0)\nonumber\\
%(m,l)=(1,2):&& \partial^3g_{x_0}(x_0)\partial u(x_0),\ \ \partial^4g_{x_0}(x_0)u(x_0) \nonumber\\
%(m,l)=(2,3):&&\partial^3g_{x_0}(x_0)\partial^3g_{x_0}(x_0) u(x_0)
%\ee
%where the indices of $\partial$ are contracted with the Hessian matrix $H(x_0)$.

%\item In the case $s=2$, the possible $(m,l)$ are $(m,l)=(0,2),(1,3),(2,4),(3,5),(4,6)$ to satisfy $2l\geq 3m$. The corresponding terms are of the types
%\be
%(m,l)=(0,2):&& \partial^4u(x_0)\nonumber\\
%(m,l)=(1,3):&& \partial^pg_{x_0}(x_0)\partial^q u(x_0),\ \ \ \
%(p\geq 3,\ p+q=6) \nonumber\\
%(m,l)=(2,4):&& \partial^{p_1}g_{x_0}(x_0)\partial^{p_2}g_{x_0}(x_0) \partial^qu(x_0)\ \ \ \
%(p_1,p_2\geq 3.\ p_1+p_2+q=8)\nonumber\\
%(m,l)=(3,5):&& \partial^{p_1}g_{x_0}(x_0)\partial^{p_2}g_{x_0}(x_0)\partial^{p_3}g_{x_0}(x_0) \partial^qu(x_0)\ \ \ \  (p_1,p_2,p_3\geq 3.\ p_1+p_2+p_3+q=10)\nonumber\\
%(m,l)=(4,6):&& \partial^{3}g_{x_0}(x_0)\partial^{3}g_{x_0}(x_0)\partial^{3}g_{x_0}(x_0)\partial^{3}g_{x_0}(x_0) u(x_0)
%\ee
%where the indices of $\partial$ are contracted with the Hessian matrix $H(x_0)$.

%\end{itemize}

We apply the above general result to the path integral representation of $A_{\l j_f}(\ck)$ restricting into a compact neighborhood of the background $\background$. The full integral of $A_{\l j_f}(\ck)$ may be understood as an (infinite) sum of the integrals over compact neighborhoods. On each of the compact neighborhood, if there is a critical point, we can employ Theorem \ref{asymptotics}(A) to determine the asymptotic expansion of the partial amplitude $A_{\l j_f}(\ck)$. In case there is no critical point in the neighborhood, the partial amplitude decays faster than $(1/\l)^k$ for all $k\in\Z_+$, provided that $\sup{\lt|D^\a u\rt|}{[|S'|^2+\Re(S)]^{-k+|\a|/2}}$ is finite (i.e. doesn't cancel the $(1/\l)^k$ behavior in front).

It is clear that the action $S[j_f,g_{ve},z_{vf}]$ of the path integral representation of the partial amplitude $A_{\l j_f}(\ck)$ depends on spin parameters $j_f$, which are not integrated in $A_{\l j_f}(\ck)$. We will see in the following that some values of $j_f$'s doesn't result in solution of the critical equations $S'=0$ and $\Re(S)=0$, which has interpretation of nondegenerate geometry. Here we call these values of $j_f$ \emph{Non-Regge-like}, otherwise we call them \emph{Regge-like}. A Regge-like spin configuration $j_f$ admits the critical configuration $(j_f,g_{ve},z_{vf})$ which can be interpreted as nondegenerate geometry. We consider a neighborhood $\O$\footnote{$\O$ is a compact neighborhood in the space of spinfoam configurations $\cm_{j,g,z}=\R_+^{\#_f}\times \Slc^{\#_{(e,v)}}\times (\mathbb{CP}^1)^{\#_{v,f}}$ modulo gauge transformations. It can be viewed as that we choose the test function $u(x)$ in Theorem \ref{asymptotics} to be compact support on a neighborhood $K\subset\Slc^{\#_{(e,v)}}\times (\mathbb{CP}^1)^{\#_{v,f}}$, then $\O=[0,\L]^{\#_f}\times K$ where $\L$ is the cut-off of the spins.} of a critical configuration $(j_f,g_{ve},z_{vf})$ with nondegenerate geometrical interpretation. Assuming this neighborhood $\O$ doesn't touch any critical configuration of degenerate geometry\footnote{The degeneracy is a special restriction Eq.\eqref{proddet} of the group variables $g_{ve}$ into a lower-dimensional submanifold. Thus such a neighborhood always exists. See the discussion about the geometrical interpretation of critical configurations for details.}, the Non-Regge-like spin configurations in $\O$ doesn't result in any critical point. For these Non-Regge-like spins, however, it is \emph{not} necessary that the corresponding $A_{\l j_f}(\ck)$, when the integral is restricted in $\O$, decays faster than $(1/\l)^k$ for all $k\in\Z_+$. As an example, for a non-Regge-like $j_f$ infinitesimally close to a Regge-like $j_f$ (the minimal gap $\Delta j_f=\frac{1}{2\l}$), $\sup{\lt|D^\a u\rt|}{[|S'|^2+\Re(S)]^{-k+|\a|/2}}$ is likely to be large and cancel the $(1/\l)^k$ behavior. Therefore if we consider the semiclassical behavior of the state-sum $A(\ck)=\sum_{j_f}\mu(j_f)A_{j_f}(\ck)$ in the large spin regime, we should in general not ignore these ``almost Regge-like'' contributions. Here in this paper we consider mainly the asymptotic behavior of $A_{\l j_f}(\ck)$ with Regge-like $j_f$, when we discuss the nondegenerate geometrical interpretation. A perturbative analysis of the semiclassical spinfoam state-sum including all possible $j_f$'s is presented in a companion paper \cite{statesum}.

\section{Bivector Interpretation of Critical Configurations}\label{bivector}

It is clear from Theorem \ref{asymptotics} that the leading order asymptotics of the spinfoam amplitude $A_{\l j}(\ck)$ is controlled by the solutions of critical equations $S'=0$ and $\Re(S)=0$, where $S$ is the spinfoam action defined in Eq.\eqref{SSS}. It turns out that $S'=0$ and $\Re(S)=0$ are equivalent to the following equations. See Appendix \ref{CE} for the derivations.
\be
\Re(S)=0:&& \frac{g^\dagger_{ve}z_{vf}}{\lt|\lt|Z_{vef}\rt|\rt|}=e^{i\a^f_{vv'}}\frac{g^\dagger_{v'e}z_{v'f}}{\lt|\lt|Z_{v'ef}\rt|\rt|},\label{1}\\
\delta_{z_{vf}}S=0:&& \frac{g_{ve}g_{ve}^\dagger z_{vf}}{\lag Z_{vef},Z_{vef}\rag}=\frac{g_{ve'}g_{ve'}^\dagger z_{vf}}{\lag Z_{ve'f},Z_{ve'f}\rag},\label{2}\\
\delta_{g_{ve}}S=0:&& \sum_{f}j_{f}\eps_{ef}(v)\frac{ \lag Z_{vef}\ \vec{\sig}\ Z_{vef}\rag }
{\lag Z_{vef},Z_{vef}\rag}=0.\label{3}
\ee
The solutions of the critical equations have interesting geometrical interpretation. In order to explain the geometrical interpretation of the solutions, it is useful to introduce a set of new variables and their relations, which mediates the spinfoam variables and the variables describing geometry. 

First of all, Eq.\eqref{1} motivates us to define an auxiliary normalized spinor $\zeta_{ef}$ by
\be
\zeta_{ef}=e^{i\phi^f_{ev}}\frac{g^\dagger_{ve}z_{vf}}{\lt|\lt|Z_{vef}\rt|\rt|}=e^{i\phi^f_{ev'}}\frac{g^\dagger_{v'e}z_{v'f}}{\lt|\lt|Z_{v'ef}\rt|\rt|},\ \ \ \ \phi^f_{ev'}-\phi^f_{ev}=\a_{vv'}^f\label{zeta}
\ee 
It is clear that the auxiliary spinor $\zeta_{ef}$ is defined up to a U(1) gauge transformation $\zeta_{ef}\mapsto e^{i\theta_{ef}}\zeta_{ef}$ or $\phi^f_{ev}\mapsto \phi^f_{ev}-\theta_{ef}$, which leaves $\a_{vv'}^f$ thus Eq.\eqref{1} invariant. From this definition, we have for each vertex $v$ connecting 2 edges $e,e'\subset\partial f$
\be
\left( g_{ve}^{\dag }\right) ^{-1}\zeta_{ef}=\frac{\left\Vert Z_{ve^{\prime
}f}\right\Vert }{\left\Vert Z_{vef}\right\Vert }e^{i\phi _{eve^{\prime
}}}\left( g_{ve^{\prime }}^{\dag }\right) ^{-1}\zeta_{e^{\prime }f}
\ee
where $\phi_{eve'}=\phi^f_{ev}-\phi^f_{e'v}$ is invariant under the U(1) gauge transformation of the auxiliary spinor $\zeta_{ef}$. By the anti-linear operator $J$ defined by $J(z_0,z_1)^t=(-\bar{z}_1,\bar{z}_0)^t$, we have the relation $JgJ^{-1}=(g^\dagger)^{-1}$ for any $g\in\Slc$. Then the above equation is equivalent to
\be
 g_{ve}(J\zeta_{ef})=\frac{\left\Vert Z_{ve^{\prime
}f}\right\Vert }{\left\Vert Z_{vef}\right\Vert }e^{-i\phi _{eve^{\prime
}}} g_{ve^{\prime }} (J\zeta_{e^{\prime }f})\label{zeta1}
\ee
This equation is equivalent to Eq.\eqref{1} i.e. $\Re(S)=0$. 

By the auxiliary spinor $\zeta_{ef}$, Eq.\eqref{2} i.e. $\delta_{z_{vf}}S=0$, is equivalent to
\be
{g_{ve}\zeta_{ef}}=\frac{|| Z_{vef}||}{|| Z_{ve'f}||}e^{i\phi_{eve'}}g_{ve'}\zeta_{e'f}. \label{zeta2}
\ee

Finally the closure condition Eq.\eqref{3} is equivalent to
\be
\sum_{f}\eps_{ef}(v)j_{f}{ \lag \zeta_{ef}\ \vec{\sig}\ \zeta_{ef}\rag}\equiv \sum_{f}\eps_{ef}(v)j_{f}\hat{n}_{ef}
=0\label{zeta3}
\ee
where $\hat{n}_{ef}\equiv \hat{n}_{\zeta_{ef}}$ and 
\be
\hat{n}_\zeta=(\zeta^0\bar{\zeta}^1+\zeta^1\bar{\zeta}^0)\hat{\mathbf{x}}-i(\zeta^0\bar{\zeta}^1-\zeta^1\bar{\zeta}^0)\hat{\mathbf{y}}+(\zeta^0\bar{\zeta}^0-\zeta^1\bar{\zeta}^1)\hat{\mathbf{z}}
\ee
is a unit 3-vector since $\zeta_{ef}$ is a normalized spinor.

Introducing the auxiliary spinor $\zeta_{ef}$ results in a set of equations Eqs.\eqref{zeta1}, \eqref{zeta2}, and \eqref{zeta3} which is equivalent to the set of critical equations Eqs.\eqref{1}, \eqref{2}, and \eqref{3}. Thus a solution $(j_f,g_{ve},z_{vf})$ of the critical equations is 1-to-1 corresponding to a solution $(j_f,g_{ve},z_{vf},\zeta_{ef})$ of Eqs.\eqref{zeta1}, \eqref{zeta2}, and \eqref{zeta3}, with $\zeta_{ef}$ determined by Eq.\eqref{zeta} up to a phase. However Eqs.\eqref{zeta1}, \eqref{zeta2}, and \eqref{zeta3} are identical to the critical equations derived in \cite{semiclassical,HZ} from a different spinfoam action.

As it was discussed previously, given a normalized 2-spinor $\zeta$, it naturally constructs a null vector $\iota(\zeta)$ via $\zeta{\zeta}^{\dagger}=\frac{1}{\sqrt{2}}\iota(\zeta)^I\sig_I$ where $\sig_I=(1,\vec{\sig})$. It is straight-forward to check that
\be
\zeta{\zeta}^\dagger=\half(1+\vec{\sig}\cdot\hat{n}_\zeta)\ \ \ \ \
\text{with}\ \ \ \ \
\hat{n}_\zeta=(\zeta^0\bar{\zeta}^1+\zeta^1\bar{\zeta}^0)\hat{\mathbf{x}}-i(\zeta^0\bar{\zeta}^1-\zeta^1\bar{\zeta}^0)\hat{\mathbf{y}}+(\zeta^0\bar{\zeta}^0-\zeta^1\bar{\zeta}^1)\hat{\mathbf{z}}\label{zetazeta}
\ee
$\hat{n}_\zeta$ is a unit 3-vector since $\zeta$ is a normalized spinor. Thus we obtain that
\be
\iota(\zeta)=\frac{1}{\sqrt{2}}(1,\hat{n}_\zeta)
\ee
Similarly for the spinor $J\zeta$, we define the null vector $(J\zeta)(J\zeta)^{\dagger}=\frac{1}{\sqrt{2}}\iota(J\zeta)^I\sig_I$ and obtain that
\be
\iota(J{\zeta})=\frac{1}{\sqrt{2}}(1,-\hat{n}_\zeta)
\ee
It is clear that $\iota({\zeta}),\iota(J{\zeta})$ is independent of the U(1) gauge freedom of $\zeta$. We can write Eqs.\eqref{zeta1} and \eqref{zeta2} in their vector representation
\be
{g}_{ve}\act \iota( J\zeta_{ef})  =\frac{\left\Vert Z_{ve^{\prime
}f}\right\Vert^2 }{\left\Vert Z_{vef}\right\Vert^2 }{g}_{ve^{\prime }}\act \iota( J\zeta_{e^{\prime }f})
\ \ \ \ \text{and}\ \ \ \
{g}_{ve}\act \iota(\zeta_{ef}) =\frac{\left\Vert Z_{vef}\right\Vert ^2}{\left\Vert
Z_{ve^{\prime }f}\right\Vert^2 }{g}_{ve^{\prime
}}\act \iota(\zeta_{e^{\prime }f})
\ee
where $g\act \iota(\zeta)$ is obtained via $g\act \zeta\zeta^\dagger=g\zeta\zeta^\dagger g^\dagger$.

It is obvious that if we construct a bivector\footnote{the pre-factor is a convention for simplifying the notation in the following discussion.}
\be
X_{ef}^{IJ}=-2\g j_f\lt[\iota(\zeta_{ef})\wedge\iota(J\zeta_{ef})\rt]^{IJ}\label{Xef0}
\ee
$X_{ef}$ satisfies the parallel transportation condition within a 4-simplex
\be
({g}_{ve})^{I}_{\ K}({g}_{ve})^{J}_{\ L}X^{KL}_{ef}=({g}_{ve'})^{I}_{\ K}({g}_{ve'})^{J}_{\ L}X^{KL}_{e'f}.
\ee
We define the bivector $X^{IJ}_f$ located at each vertex $v$ of the dual face $f$ by the parallel transportation
\be
X_f^{IJ}(v):=({g}_{ve})^{I}_{\ K}({g}_{ve})^{J}_{\ L}X^{KL}_{ef}.
\ee
which is independent of the choice of $e$ by the above parallel transportation condition. Then we also have the parallel transportation relation of $X^{IJ}_f(v)$
\be
X_f^{IJ}(v)=({g}_{vv'})^{I}_{\ K}({g}_{vv'})^{J}_{\ L}X^{KL}_{f}(v').
\ee
where $g_{vv'}=g_{ve}g_{ev'}$ with $g_{ev}=g_{ve}^{-1}$.

On the other hand, we can write the bivector $X^{IJ}_{ef}$
\be
X_{ef}\equiv (X_{ef})^{I}_{\ J}=2\g j_f
\left(\begin{array}{cccc}0 &\ \hat{n}^1_{ef} &\ \hat{n}^2_{ef} &\ \hat{n}^3_{ef} \\ \hat{n}^1_{ef} &\ 0 &\ 0 &\ 0 \\ \hat{n}_{ef}^2 &\ 0 &\ 0 &\ 0 \\  \hat{n}^3_{ef} &\ 0 &\ 0 &\ 0\end{array}\right)=2\g j_f\hat{n}_{ef}\cdot\vec{K}\label{Xef}
\ee
where $\vec{K}$ denotes the boost generator of Lorentz Lie algebra $\slc$ in the bivector representation. The rotation generator in $\slc$ is denoted by $\vec{J}$. The generators in $\slc$ satisfies the commutation relations $[J^i,J^j]=-\eps^{ijk}J^k,\ [J^i,K^j]=-\eps^{ijk}K^k,\ [K^i,K^j]=\eps^{ijk}J^k$. The relation $X_{ef}=2\g j_f\hat{n}_{ef}\cdot\vec{K}$ gives a representation of the bivector in terms of the $\slc$ lie algebra generators. Moreover it is not difficult to verify that in the spin-$\half$ representation $\vec{J}=\frac{i}{2}\vec{\sig}$ and $\vec{K}=\frac{1}{2}\vec{\sig}$. Thus in the spinor representation
\be
X_{ef}= \g j_f\vec{\sig}\cdot\hat{n}_{ef}\label{Xsig}
\ee
For this $\slc$ Lie algebra representation of the bivector $X_{ef}$, the parallel transportation is represented by the adjoint action of the Lie group on its Lie algebra. Therefore we have
\be
g_{ve}X_{ef}g_{ev}=g_{ve'}X_{e'f}g_{e'v},\ \ \ \ X_{f}(v):=g_{ve}X_{ef}g_{ev},\ \ \ \ X_{f}(v):=g_{vv'}X_{f}(v')g_{v'v}
\ee
where $g_{ve}=g_{ev}^{-1}, g_{v'v}=g_{vv'}^{-1}$. We note that the above equations are valid for all the representations of $\slc$.

There is a duality map acting on $\slc$ by $*\vec{J}=-\vec{K}$, $*\vec{K}=\vec{J}$ with $*^2=-1$. For self-dual/anti-self-dual bivector $\vec{T}_\pm:=\half(\vec{J}\pm i\vec{K})$, One can verify that $*\vec{T}_\pm=\pm i\vec{T}_\pm$. In the bivector representation, the duality map is represented by $*X^{IJ}=\half\eps^{IJKL}X_{KL}$. In the spinor representation, the duality map is represented by $*X=iX$ since $\vec{J}=\frac{i}{2}\vec{\sig}$ and $\vec{K}=\frac{1}{2}\vec{\sig}$ in the spinor representation. From Eq.\eqref{Xef}, we see that
\be
X_{ef}=-*(2\g j_f\hat{n}_{ef}\cdot\vec{J})
\ee
From its bivector representation one can see that
\be
\eta_{IJ}u^I*\!X_{ef}^{JK}=0,\ \ \ \ \ u^I=(1,0,0,0).
\ee
It motivates us to define a unit vector at each vertex $v$ for each tetrahedron $t_e$ by
\be
N^I_e(v):=({g}_{ve})^{I}_{\ J}u^J\label{Ne}
\ee
Then for all triangles $f$ in the tetrahedron $t_e$, $N^I_e(v)$ is orthogonal to all the bivectors $*\!X_f(v)$ with $f$ belonging to $t_e$.
\be
\eta_{IJ}N^I_e(v)*\!X^{JK}_f(v)=0.
\ee

In addition, from the closure constraint Eq.\eqref{closure}, we obtain for each tetrahedron $t_e$
\be
\sum_{f\subset t_e}\eps_{ef}(v)X_f(v)=0.
\ee

So far the bivectors $X_{ef}$ and $X_f(v)$ are expressed in terms of $j_f,\ g_{ve}$ and the auxiliary spinor $\zeta_{ef}$. However they can be expressed in terms of $j_f,\ g_{ve}$ and $z_{vf}$ by Eq.\eqref{zeta}. Firstly by definition
\be
\zeta_{ef}\zeta_{ef}^\dagger&=&\frac{g_{ve}^\dagger z_{vf}z_{vf}^\dagger g_{ve}}{\lag g_{ve}^\dagger z_{vf},g_{ve}^\dagger z_{vf}\rag}=\frac{1+g^\dagger_{ve}(\hat{n}_{vf})\cdot\vec{\sig}}{2}\nonumber\\
(J\zeta_{ef})(J\zeta_{ef})^\dagger&=&\frac{g_{ve}^{-1} (Jz_{vf})(Jz_{vf})^\dagger (g_{ve}^{-1})^\dagger}{\lag g_{ve}^\dagger z_{vf},g_{ve}^\dagger z_{vf}\rag}=\frac{1-g^{-1}_{ve}(\hat{n}_{vf})\cdot\vec{\sig}}{2}
\ee
where $\hat{n}_{vf}\in S^2$ is given by $\hat{n}_{vf}=\langle z_{vf}\ \vec{\sig}\ z_{vf}\rangle$, and $g_{ve}(\hat{n}_{vf})$ is the representation of $\Slc$ on 2-sphere, which shouldn't be confused with the vector representation of Lorentz group. Comparing the above equation with Eq.\eqref{zetazeta}, we have the relation
\be
g^\dagger_{ve}(\hat{n}_{vf})=\hat{n}_{ef}.
\ee
Thus the bivector $X_{ef}$ an be written as
\be
X^{IJ}_{ef}=2\g j_f\lt[u\wedge\hat{n}_{ef}\rt]^{IJ}=2\g j_f\lt[u\wedge g^\dagger_{ve}(\hat{n}_{vf})\rt]^{IJ},\ \ \ \ u=(1,0,0,0)^t
\ee
On the other hand, if we write $X_{ef}=2\g j_f\lt[\iota(J\zeta_{ef})\wedge\iota(\zeta_{ef})\rt]$ or we can denote
\be
X_{ef}=2\g j_f\lt[(J\zeta_{ef})(J\zeta_{ef})^\dagger\wedge \zeta_{ef}\zeta_{ef}^\dagger\rt]= 2\g j_f\lt[g_{ve}^{-1} (Jz_{vf})(Jz_{vf})^\dagger (g_{ve}^{-1})^\dagger\wedge \frac{g_{ve}^\dagger z_{vf}z_{vf}^\dagger g_{ve}}{\lag g_{ve}^\dagger z_{vf},g_{ve}^\dagger z_{vf}\rag^2}\rt]
\ee
Then $X_f(v)=g_{ve}\act X_{ef}$ is given by 
\be
X_f(v)=2\g j_f\lt[(Jz_{vf})(Jz_{vf})^\dagger\wedge \frac{g_{ve}g_{ve}^\dagger z_{vf}z_{vf}^\dagger g_{ve}g_{ve}^\dagger}{\lag g_{ve}^\dagger z_{vf},g_{ve}^\dagger z_{vf}\rag^2}\rt]
\ee
which is clearly independent of the choice of $e$ and $e'$ connecting to $v$ by the equation of motion Eq.\eqref{2}.

To summarize, the above analysis shows the following result: Given the data $(j_f,g_{ev},z_{vf})$ be a spinfoam critical configuration that solves the critical equations Eqs.\eqref{1}, \eqref{2}, and \eqref{3}, the data determine uniquely a set of bivector variables (in the $\slc$ Lie algebra representation) $X_{ef}(v)$ with $|X_{ef}(v)|=\sqrt{\half\tr\lt(X_{ef}(v)X_{ef}(v)\rt)}=2\g j_f$, which satisfies
\be
&&X_{ef}(v)=X_{e'f}(v)\equiv X_f(v),\ \ \ \ X_{f}(v):=g_{vv'}X_{f}(v')g_{v'v}, \nonumber\\
&&\eta_{IJ}N^I_e(v)*\!X^{JK}_f(v)=0,\ \ \ \ \ \ \ \sum_{f\subset t_e}\eps_{ef}(v)X_f(v)=0.\label{summarize}
\ee
where $e$ and ${e'}$ are two edges connecting to $v$, $f$ is the dual face determined by $v, e,e'$, and $N^I_e(v)=({g}_{ve})^{I}_{\ J}u^J$ with $u^J=(1,0,0,0)$ is a unit vector. Such a result obtained in the above analysis coincides precisely with the bivector interpretation of critical configuration obtained in \cite{HZ} (Proposition 2.1). Therefore the geometrical interpretation of the critical configuration $(j_f,g_{ev},z_{vf})$ can be analyzed in the same way as it was done in \cite{HZ}. We reproduce the analysis in the next section, in order to set up the notion for the discussion in Sections \ref{GeoEq}, \ref{SG}, and \ref{timeorientation}.

\section{Nondegenerate Lorentzian Geometrical Interpretation}

Given a spinfoam critical configuration $(j_f,g_{ev},z_{vf})$ that solves the critical equations, let's consider a triangle $f$ shared by two tetrahedra $t_e$ and $t_{e'}$ of a 4-simplex $\sig_v$. There are the simplicity conditions $N^e_I(v) *\!X^{IJ}_{ef}(v)=0$ and $N^{e'}_I(v) *\!X^{IJ}_{e'f}(v)=0$ from the viewpoint of the two tetrahedra $t_e$ and $t_{e'}$. Recall that $N_e(v)=g_{ve}\act (1,0,0,0)^t$. The two simplicity conditions imply that there exists two 4-vectors $M_{ef}^I(v)$ and $M_{ef}^I(v)$ such that $X_{ef}(v)=N_e(v)\wedge M_{ef}(v)$ and $X_{e'f}(v)=N_{e'}(v)\wedge M_{e'f}(v)$. However we have the gluing condition $X_{ef}(v)=X_{e'f}(v)=X_{f}(v)$, which implies that $N_{e'}(v)$ belongs to the plane spanned by $N_e(v), M_{ef}(v)$, i.e. $N_{e'}(v)=a_{ef}M_{ef}(v)+b_{ef}N_e(v)$. In this section we assume the following nondegeneracy condition involving purely the group variables \cite{HZ,CF}: For all the choices of four different edges $e_1,e_2,e_3,e_4=1,\cdots,5$ 
\be
\det\Big(N_{e_1}(v),N_{e_2}(v),N_{e_3}(v),N_{e_4}(v)\Big)\neq0,\ \ \ \forall\ v\label{proddet}
\ee
then $N_e(v),N_{e'}(v)$ cannot be parallel to each other, for all pairs of $e,e'$, which excludes the case of vanishing $a_{ef}$ in the above. Denoting $\a_{ee'}=a_{ef}^{-1}$, we obtain that $M_{ef}(v)=\a_{ee'}N_{e'}(v)-\a_{ee'}b_{ef}N_e(v)$. Therefore
\be
X_{f}(v)\equiv X_{ee'}(v)=\a_{ee'}(v)\lt[N_{e}(v)\wedge N_{e'}(v)\rt]\label{NN}
\ee
for all $f$ shared by $t_e$ and $t_{e'}$. If we label the 5 tetrahedra of $\sig_v$ by $t_{e_i}$, $i=1,\cdots,5$. Then Eq.\eqref{NN} can be written as $X_{e_i e_j}(v)=\a_{ij}(v)\lt[N_{e_i}(v)\wedge N_{e_j}(v)\rt]$. Then the closure condition $\sum_{j=1}^4\eps_{e_i e_j}(v)X_{e_i e_j}(v)=0$\footnote{Here $\eps_{e_ie_j}(v)=-\eps_{e_je_i}(v)$ and $X_{e_i e_j}(v)=X_{e_j e_i}(v)$.} gives that $\forall\ i=1,\cdots,5$
%\footnote{The sign factor $\eps_{e_i e_j}(v)=\eps_{e_i f}$ depends only on whether the orientations of the edge $e$ and the dual face $f$ are consistent or not, thus it doesn't depend on the vertex $v$. However the dependence of $v$ in the notation $\eps_{e_i e_j}(v)$ is because the dual face $f$ is determined by a triple $(v,e_i,e_j)$. }
\be
0=\sum_{j=1}^4\eps_{e_i e_j}(v)\a_{ij}(v)\lt[N_{e_i}(v)\wedge N_{e_j}(v)\rt]=N_{e_i}(v)\wedge \sum_{j=1}^4\eps_{e_i e_j}(v)\a_{ij}(v)N_{e_j}(v)
\ee
which implies that for a choice of diagonal element $\b_{ii}(v)$,
\be
\sum_{j=1}^5\b_{ij}(v)N_{e_j}(v)=0\label{sumN}
\ee
where we denote $\b_{ij}(v):=\eps_{e_i e_j}(v)\a_{ij}(v)$. Here $\b_{ii}(v)$ must be chosen as nonzero, because if $\b_{ii}(v)=0$, Eq.\eqref{sumN} would reduce to $\sum_{j\neq i}\b_{ij}(v)N_{e_j}(v)=0$, which gives all the coefficients $\b_{ij}(v)=0$ by linearly independence of any four $N_e(v)$ (from the nondegeneracy Eq.\eqref{proddet}).

\begin{Lemma}\label{factorization}

Assuming the nondegeneray condition Eq.\eqref{proddet}, Eq.\eqref{sumN} implies a factorization of $\b_{ij}(v)$
\be
\b_{ij}(v)=\varsigma(v)\b_i(v)\b_j(v)
\ee
where $\varsigma(v)=\pm1$. Thus we have the following expression of the bivector $\eps_{e_i e_j}(v)X_{e_i e_j}(v)$
\be
\eps_{e_i e_j}(v)X_{e_i e_j}(v)=\tilde{\eps}(v)\Big(\b_i(v)N_{e_i}(v)\Big)\wedge \Big(\b_j(v)N_{e_j}(v)\Big)
\ee
Eq.\eqref{sumN} then takes the form
\be
\sum_{j=1}^5\b_{j}(v)N_{e_j}(v)=0.\label{bN}
\ee
$\b_i(v)$ satisfies the following relation with $j_f$ and $g_{ve}$
\be
4\g^2 j_f^2=\b_e(v)^2\b_{e'}(v)^2\lt[1-\cosh^2\theta_{ee'}(v)\rt],\ \ \ \ \cosh\theta_{ee'}(v)=N_e(v)\cdot N_{e'}(v).\label{jbb}
\ee

\end{Lemma}

See Appendix \ref{5.1} for a proof.

Now we construct 5 vectors $U_{e_i}(v)$ at each vertex $v$\footnote{We denote the dual vector $N^e_I$ by $N^e$ and the vector $N_e^I$ by $N_e$, and the same convention holds for $U_e$ and $U^e$. } up to a sign $\pm_v$ globally for all the 5 $U_{e}(v)$ at $v$:
\be
U^{e_i}_I(v):=\pm_v\frac{\b_i(v)N^{e_i}_I(v)}{\sqrt{|V_4(v)|}}\ \ \text{with}\ \ V_4(v):=\det\Big(\b_2(v)N^{e_2}(v),\b_3(v)N^{e_2}(v),\b_4(v)N^{e_2}(v),\b_5(v)N^{e_2}(v)\Big)\label{U}
\ee
Any four of the five  $U_{e_i}(v)$ span a 4-dimensional vector space by the assumption of nondegeneracy. Although $N_e(v)$ is always future-pointing since $N_{e}(v)=g_{ve}(1,0,0,0)^t$ where$\ g_{ve}\in\Slc$, the vectors $U_e(v)$ is future-pointing if $\pm_v\b_e(v)>0$, and is past-pointing if $\pm_v\b_e(v)<0$. Moreover $U_e(v)$'s satisfy
\be
\sum_{j=1}^5U_{e_j}(v)=0\ \ \ \ \text{and}\ \ \ \
\frac{1}{V_4(v)}=\det\Big(U^{e_2}(v),U^{e_3}(v), U^{e_4}(v),U^{e_5}(v)\Big)
\ee
where the 4-simplex closure condition implies that there is at least one future-pointing $U_e(v)$ and one past-pointing $U_e(v)$ at each $v$. Now the bivectors can be expressed as
\be
\eps_{e_i e_j}(v)X^{e_i e_j}_{IJ}(v)=\varsigma(v)|V_4(v)|\lt[ U_{e_i}(v)\wedge U_{e_j}(v)\rt]_{IJ}=\eps(v)V_4(v)\lt[ U_{e_i}(v)\wedge U_{e_j}(v)\rt]_{IJ}
\ee
where $\eps(v)=\varsigma(v)\mathrm{sgn}(V_4(v))$. The oriented 4-volume $V_4(v)$ in general can be either positive or negative for different 4-simplices. %However if we consider the perturbations around a background configuration of $V_4(v)>0$, then the perturbative corrections never change the sign of $\mathrm{sgn}(V_4(v))$.

We construct the inverse $E_{e_i e_j}(v)$ $i,j=1,\cdots,5$ as a collection of space-like 4-vectors, such that
\be
U^{e_i}_I(v) E^I_{e_j e_k}(v)=\delta^i_j-\delta^i_k\ \ \ \ \text{and}\ \ \ \ E_{e_i e_j}^{I}(v)=-E_{e_j e_i}^{I}(v)
\ee
One can also verify immediately the following ``triangle closure condition''
\be
E^I_{e_j e_k}(v)+E^I_{e_k e_l}(v)+E^I_{e_l e_j}(v)=0
\ee
Inside a 4-simplex $\sig_v$ dual to $v$, a set of 10 vectors $E_{e_i e_j}(v)$ $i,j=1,\cdots,5$ satisfies the above triangle closure condition for each triple $(j,k,l)$ has a geometrical interpretation as a set of edge-vectors of the 4-simplex. An edge $\ell$ can be denoted by its end-points, say $p_1,p_2$, i.e. $\ell=[p_1,p_2]$. There are 5 vertices $p_i,i=1,\cdots,5$ for a 4-simplex $\sig_v$. Then each $p_i$ is one-to-one corresponding to a tetrahedron $t_{e_i}$ of the 4-simplex $\sig_v$. Therefore we can denote the oriented edge $\ell=[p_1,p_2]$ also by $\ell=(e_1,e_2)$, once a 4-simplex $\sig_v$ is specified. Thus we denote by $E_{\ell}(v)\equiv E_{e_1e_2}(v)$ the edge-vector associated with the edge $\ell=(e_1,e_2)$. The norm of the edge-vector $E_{\ell}(v)$ is the edge-length $s_\ell:=|E_{\ell}(v)|$. Thus we have shown that for each individual 4-simplex, there exists a set of edge-vector $E_{\ell}(v)$ with the edge-lengths $s_\ell$ constructed from the spinfoam critical configuration $(j_f,g_{ve},z_{vf})$.

In the context of simplical geometry \cite{deficit}, the collection of ten $E_{ee'}(v)$'s is called a discrete cotetrad at the vertex $v$. We have shown that given a spinfoam critical configuration $(j_f,g_{ve},z_{vf})$, there exists a discrete cotetrad $E_{ee'}(v)$ at each vertex $v$, such that the bivector $X_{ee'}(v)$ can now be expressed by $E_{ee'}(v)$
\be
\eps_{e_4 e_5}(v)X^{IJ}_{e_4e_5}(v)=\eps(v)*\! \Big[E_{e_1e_2}(v)\wedge E_{e_2 e_3}(v)\Big]^{IJ}\label{bivectorcotetrad}
\ee
which will also be denoted by $X_{f}(v)=\eps(v)*\! \Big[E_{\ell_1}(v)\wedge E_{\ell_2}(v)\Big]$ by assuming the orientations of $\ell_1,\ell_2$ is compatible with the orientation of $f$, so that the sign $\eps_{ee'}(v)$ doesn't appear. If we take the norm of the bivector, we obtain
\be
(2\g j_f)^2=s_\ell^2 s_{\ell'}^2 \sin^2\Phi_{\ell,\ell'},\ \ \ \text{where}\ \  \cos\Phi_{\ell,\ell'}=E_\ell(v)\cdot E_{\ell'}(v)/(s_\ell s_{\ell'})
\ee
By the above relation, we interpret $\g j_f$ of the critical data $(j_f,g_{ve},z_{vf})$ as a triangle area related to the 2 edge-lengths $s_{\ell},s_{\ell'}$ and the angle $\Phi_{\ell,\ell'}$ between $\ell,\ell'$. 

So far we have shown the existence of a cotetrad $E_\ell(v)$ characterizing the 4-simplex geometry from the spinfoam critical data $(j_f,g_{ve},z_{vf})$. In order to show the uniqueness of $E_\ell(v)$, we construct the 3-d angles $\Psi_{f,f'}(v)\in[0,\pi]$ uniquely from the bivectors:
\be
\cos\Psi_{f,f'}(v):=\frac{\lt[\eps_{ef}(v)*\!X_f(v)\rt]^{IJ}\lt[\eps_{ef}(v)*\!X_{f'}(v)\rt]_{IJ}}{4\g^2 j_f j_{f'}}
\ee
By the existence of edge-vectors $E_\ell(v)$ and interpretation of $\g j_f$ as the area of the triangle $f$, the angle $\Psi_{f,f'}(v)$ is the 3-d dihedral angle between the triangles $f,f'$ in the tetrahedron $t_e$, which satisfies the gluing constraint and closure constraint in the Area-Angle Regge calculus \cite{areaangle}. If we compute the 2-d angle $\Phi_{\ell,\ell'}$ in tetrahedron $t_{e_k}$ by (We follow the convention in \cite{areaangle} in this formula\footnote{$\Psi_{jk,i}$ is the 3-d angle determined by the bivectors $X_{e_ie_j}(v)$ and $X_{e_ie_k}(v)$. $\Phi_{ij,kl}$ is the 2-d angle between the edges $(p_i,p_j,p_k)^c$ and $(p_i,p_j,p_l)^c$, where e.g. $(p_1,p_2,p_3)^c$ denotes the edge $(p_4,p_5)$.}) 
\be
\cos\Phi_{ij,kl}=\frac{\cos\Psi_{ij,k}+\cos\Psi_{il,k}\cos\Psi_{jl,k}}{\sin\Psi_{il,k}\sin\Psi_{jl,k}}
\ee
The gluing constraint in area-angle Regge calculus requires the result should be independent of the two choice of tetrahedra sharing the triangle, i.e. $\Phi_{ij,kl}=\Phi_{ij,lk}=\Phi_{\ell,\ell'}$. Such a constraint is satisfied here by the existence of edge-vectors $E_\ell(v)$. On the other hand, the area-angle Regge calculus closure constraint is manifestly satisfied here by the closure condition of bivectors. It is shown in \cite{areaangle} that given the area-angle variables $\g j_f$ and $\Psi_{f,f'}$ satisfying the gluing and closure constraint in area-angle Regge calculus\footnote{The nondegenerancy assumption is also required.}, they implies a unique set of edge-lengths $s_{\ell}$ of the 4-simplex. Since the area-angle variables $\g j_f$ and $\Psi_{f,f'}$ are uniquely determined by the spinfoam critical data $(j_f,g_{ve},z_{vf})$, the spinfoam critical data $(j_f,g_{ve},z_{vf})$ then uniquely fix a set of edge-lengths $s_{\ell}$ of the 4-simplex.

There is an important difference between critical spinfoam variables and length or area-angle varaibles in Regge calculus. That is, the spinfoam critical configuration $(j_f,g_{ve},z_{vf})$ also specifies the orientation of the geometrical 4-simplex by determining $\mathrm{sgn}(V_4(v))$ by the following relation\footnote{The orientation of a (topological) 4-simplex $\sig_v$ is represented by an ordering of its 5 vertices, i.e. a tuple $[p_1,\cdots,p_5]$. Two orientations are opposite to each other if the two orderings are related by an odd permutation, e.g. $[p_1,p_2,\cdots,p_5]=-[p_2,p_1\cdots,p_5]$. We say that two neighboring 4-simplices $\sig,\sig'$ are consistently oriented, if the orientation of their shared tetrahedron $t$ induced from $\sig$ is opposite to the orientation induced from $\sig'$. For example, $\sig=[p_1,p_2,\cdots,p_5]$ and $\sig'=-[p'_1,p_2,\cdots,p_5]$ are consistently oriented since the opposite orientations $t=\pm[p_2,\cdots,p_5]$ are induced respectively from $\sig$ and $\sig'$. The simplicial complex $\ck$ is said to be orientable if it is possible to orient consistently all pair of neighboring 4-simplices. Such a choice of consistent 4-simplex orientations is called a global orientation. We assume we define a global orientation of the triangulation $\ck$. Then for each 4-simplex $\sig_v=[p_1,p_2,\cdots,p_5]$, we define the oriented volume (assumed to be nonvanishing as the nondegeneracy)
\be
V_4(v):=\det\Big(E_{e_2e_1}(v),E_{e_3e_1}(v),E_{e_4e_1}(v),E_{e_5e_1}(v)\Big).
\ee}
\be
V_4(v)=\det\Big(E_{e_2e_1}(v),E_{e_3e_1}(v),E_{e_4e_1}(v),E_{e_5e_1}(v)\Big)=-\half\eps_{e_2e_3}(v)\eps_{e_4e_5}(v)\tr\lt[X_{e_2e_3}(v)*\!X_{e_4e_5}(v)\rt].
\ee
while the Regge calculus variables doesn't contain such an information. It is discussed in Section \ref{timeorientation} that the spinfoam critical data also partially contain information about the time-orientation of the (simplicial) spacetime.

Given the uniqueness of the set of edge-lengths $s_{\ell}$ of the 4-simplex, determined by the critical data $(j_f,g_{ve},z_{vf})$, the edge-vectors $E_\ell(v)$ of all the oriented edges $\ell$ of the 4-simplex are determined up to a global O(1,3) transformation acting simultaneously on all $E_\ell(v)$'s. 
\begin{itemize}

\item $\mathrm{SO}^+(1,3)$: A $\mathrm{SO}^+(1,3)$ transformation simultaneously acting on all $E_\ell(v)$'s results in a global $\mathrm{SO}^+(1,3)$ transformation on all $U_e(v)$'s thus $N_e(v)$'s. It corresponds to a $\Slc$ gauge transformation at each vertex $v$ acting on the critical data, i.e. $g_{ve}\mapsto x_v^\dagger g_{ve},z_{vf}\mapsto x^{-1}_v z_{vf}$ while the auxiliary spinor $\zeta_{ef}$ is invariant under such a gauge transformation. In another words, if we gauge-fix the variables $g_{ve},z_{vf}$ in the critical data, the global $\mathrm{SO}^+(1,3)$ transformation on $E_\ell(v)$'s is not allowed.

\item Parity $\mathbf{P}$: If we define ${E}_\ell(v)\mapsto\tilde{E}_\ell(v)=\mathbf{P}{E}_\ell(v)$, then $U_e(v)\mapsto \tilde{U}_e(v)=\mathbf{P}U_e(v)$, the bivector $X_f(v)\mapsto\tilde{X}_f(v)=-(\mathbf{P}\otimes\mathbf{P})X_f(v)$, and the oriented 4-volume $V_4(v)\mapsto\tilde{V}_4(v)=-V_4(v)$. Such a transformation is not allowed if we specify the critical data $(j_f,g_{ve},z_{vf})$ since the critical data specifies uniquely the bivector and the oriented 4-volume. It turns out that the parity inversion ${E}_\ell(v)\mapsto\tilde{E}_\ell(v)$ corresponds to the following transformation between 2 \emph{different} critical configurations $(j_f,g_{ve},z_{vf})\mapsto (j_f,\tilde{g}_{ve},\tilde{z}_{vf})$ with \cite{semiclassical,HZ}
\be
\tilde{g}_{ve}=Jg_{ve}J^{-1}=(g_{ve}^\dagger)^{-1},\ \ \ \tilde{z}_{vf}=\frac{g_{ve}g_{ve}^\dagger z_{vf}}{||g_{ve}^\dagger z_{vf}||^2}
\ee 
where on the right hand side of the second relation $z_{vf}$ is assumed to satisfy the critical equations, especially Eq.\eqref{2}. We call such a transformation a ``solution-generating map'' since it generates a new solution of critical equations from a given solution. We discuss in detail the solution-generating maps in Section \ref{SG}.

\item Parity-Time inversion $\mathbf{PT}$: Globally flipping sign ${E}_\ell(v)\mapsto-_v{E}_\ell(v)$ is allowed by the critical data $(j_f,g_{ve},z_{vf})$ because by the previously discussion $U_e(v)$ and ${E}_\ell(v)$ are defined by the critical data up to $\pm_v$. 

\end{itemize}

Therefore we have shown that a spinfoam critical data $(j_f,g_{ve},z_{vf})$ uniquely specifies a discrete cotetrad $E_\ell(v)$ in each 4-simplex, up to a flip of global sign $\pm_v$. Then the 5 vectors $U_e(v)$ is also determined up to $\pm_v$ by the inverse of $E_\ell(v)$.

The above analysis concerns only a single 4-simplex $\sig_v$. The critical equation $X_f(v)=g_{vv'}X_f(v')g_{v'v}$ has not been studied yet. Now we consider two neighboring 4-simplices $\sig_v,\sig_{v'}$ with their centers $v,v'$ connected by a dual edge $e$. Here we skip the technical details of the analysis but state the result (see \cite{HZ,CF} for a proof):

\begin{Lemma}\label{globalsign}

We define a sign $\eps_e(v)=\pm1$ by $\eps_e(v)\frac{U_e(v)}{|U_e(v)|}=N_e(v)$. Given an edge $e=(v,v')$, $X_f(v)=g_{vv'}X_f(v')g_{v'v}$ and $\eps_e(v)\frac{U_e(v)}{|U_e(v)|}=g_{vv'}\eps_e(v')\frac{U_e(v')}{|U_e(v')|}$ implies that

\begin{enumerate}

\item $\eps(v)=\eps(v')=\eps$ is a global sign on the triangulation.
	
\item For all edge $\ell$ of a triangulation $t_e$ shared by the 4-simplices $\sig_v$ and $\sig_{v'}$, we have the parallel transportation relation for $E_\ell(v)$ and $E_\ell(v')$ up to a sign, i.e.
\be
\mu_e E_\ell(v)= g_{vv'}E_\ell(v')
\ee
where $\mu_e=-\eps_e(v)\eps_e(v')\mathrm{sgn}(V_4(v)V_4(v'))$. If $\O_{vv'}\in \mathrm{SO(4)}$ is the unique discrete spin connection determined by $E_\ell(v),E_\ell(v')$, then
\be
g_{vv'}=\mu_e\O_{vv'}.
\ee

\end{enumerate}
\noindent
Since $E_{\ell}(v)$ and $U_e(v)$ are determined by the critical data up to $\pm_v$, thus the sign $\eps_e(v)$ is determined up to $\pm_v$ and $\mu_e$ is determined up to $\pm_v\pm_{v'}$. The signs unambiguously determined by the critical data are $\eps_e(v)\eps_{e'}(v)$ and $\prod_{e\subset\partial f}\mu_e$.

\end{Lemma}

Assuming $(j_f,g_{ve},z_{vf})$ is a solution of the critical equations, then the parallel transportation relation of the resulting cotetrads located at different vertices implies the edge-length $s_\ell=|E_\ell(v)|=|E_{\ell}(v')|$ of an edge shared by 2-simplices $\sig_v,\sig_{v'}$ is single-valued and independent of $\sig_v$-viewpoint or $\sig_{v'}$-viewpoint. 
Therefore there exists a set of edge-lengths $s_\ell$ for all the edges $\ell$ of the simplicial complex, such that $\g j_f$ are the triangle areas consistent with the the edge-lengths. However it is well-known that such a result can only be achieved by a subclass of triangle areas \cite{arearegge}. We thus conclude that the existence of the critical configuration $(j_f,g_{ve},z_{vf})$ compatible with the nondegeneracy condition Eq.\eqref{proddet} imposes a nontrivial restriction of the spin configuration $j_f$. We call the allowed spin configurations \emph{Regge-like} spins, otherwise we call them \emph{Non-Regge-like} spins. A Regge-like spin configuration admits critical configurations with nondegenerate geometrical interpretation, while a non-Regge-like spin configuration doesn't result in any solution of the critical equation compatible with nondegeneracy condition Eq.\eqref{proddet}.

Let's summarize what we have achieved so far. The above discussion shows that a spinfoam critical configuration $(j_f,g_{ve},z_{vf})$, with a Regge-like spin configuration and satisfying the nondegeneracy assumption Eq.\eqref{proddet}, determines uniquely a set of data $(\pm_vE_\ell(v),\eps)$, where $\eps=\pm1$ is a global sign on the triangulation. The data $(\pm_vE_\ell(v),\eps)$ have the following interpretation:
\begin{itemize}

\item The set of space-like vectors $E_\ell(v)$ is a discrete cotetrad\footnote{A discrete cotetrad is a set of vectors associated with the all edges of the simplicial complex, satisfying the defining properties (all the edges $\ell$ on the triangulation are oriented)
\be
&&E_{-\ell}(v)=-E_\ell(v),\ \ \ \ E_{\ell_1}(v)+E_{\ell_2}(v)+E_{\ell_3}(v)=0,\ \ \ \ell_1,\ell_2,\ell_3\ \text{form the boundary of a triangle }f\nonumber\\
&&E_{\ell_1}(v)\cdot E_{\ell_2}(v)=E_{\ell_1}(v')\cdot E_{\ell_2}(v')\ \ \ \ \forall\ell_1,\ell_2\subset t_e\ \text{shared by $\sig_v$ and $\sig_{v'}$.}
\ee} determined up to an overall sign $\pm_v$ in each 4-simplex, such that the bivectors satisfy
\be
X_f(v)=\eps*\lt[E_{\ell_1}(v)\wedge E_{\ell_2}(v)\rt]\label{XEE}
\ee
where $\ell_1,\ell_2$ are two edges of the triangle $f$. $\eps=\pm1$ is a global sign on the entire simplical complex.

\item An oriented volume can be defined by
\be
V_4(v)=\det\lt(E_{\ell_1}(v),E_{\ell_2}(v),E_{\ell_3}(v),E_{\ell_4}(v)\rt),
\ee
and $\mathrm{sgn}(V_4(v))$ is understood as an orientation of the corresponding geometrical 4-simplex.

\item The norm of the bivector is given by $|X_f(v)|=2\g j_f$. $\g j_f$ is interpreted as the area of triangle $f$, measured in the area unit $a=\l \ell^2_P$.

\item By the inversion of the cotetrad $E_\ell(v)$ via $U^{e_i}_I(v) E^I_{e_j e_k}(v)=\delta^i_j-\delta^i_k$, we obtain 5 vectors $U_e(v)$ at each vertex $v$, which we call a discretized tetrad. We associate a sign $\eps_e(v)=\pm1$ to each vector $U_e(v)$ by requiring $\eps_e(v)=+1$ if $U_e(v)$ is future-directed and $\eps_e(v)=-1$ if $U_e(v)$ is past-directed.

\item Given an edge $e=(v,v')$ dual to a tetrahedron $t_e$, for all edges $\ell$ of the tetrahedron $t_e$
\be
\mu_e E_\ell(v)= g_{vv'}E_\ell(v').
\ee
where $\mu_e=-\eps_e(v)\eps_e(v')\mathrm{sgn}(V_4(v)V_4(v'))$. Thus a discrete metric can be constructed by
\be
\cg_{\ell_1\ell_2}(v)=E_{\ell_1}(v)\cdot E_{\ell_2}(v)
\ee
$\cg_{\ell_1\ell_2}$ is independent of the choice of $v$ or its neighbor $v'$, i.e. $\cg_{\ell_1\ell_2}(v)=\cg_{\ell_1\ell_2}(v')$ if $\ell_1,\ell_2$ are shared by the two 4-simplices. It also means that the two geometrical 4-simplices induce the same geometry on the tetrahedron they share. The discrete metric, or essentially the edge-lengths $s_\ell=\sqrt{\cg_{\ell\ell}}$, determines a discrete Lorentzian geometry on the simplicial complex. 

\end{itemize}

Note that since $E_{\ell}(v)$ and $U_e(v)$ are determined by the critical data up to $\pm_v$, the sign $\eps_e(v)$ is determined up to $\pm_v$ and $\mu_e$ is determined up to $\pm_v\pm_{v'}$. The signs unambiguously determined by the critical data are $\eps_e(v)\eps_{e'}(v)$ and $\prod_{e\subset\partial f}\mu_e$.

\section{An Equivalence Theorem}\label{GeoEq}

\begin{Theorem}\label{equivalent}

A gauge equivalence class of the spinfoam critical configuration $(j_f,g_{ve},z_{vf})$, satisfying the nondegeneracy assumption Eq.\eqref{proddet}, is equivalent to a set of geometrical data $(\pm_vE_\ell(v),\eps)$ with $V_4(v)\neq0$, where the space-like vectors $E_\ell(v)$ is a nondegenerate discrete cotetrad on the triangulation, determined up to $\pm_v$ at each $v$, satisfying the defining properties (all the edges $\ell$ on the triangulation are oriented)
\be
&&E_{-\ell}(v)=-E_\ell(v),\ \ \ \ E_{\ell_1}(v)+E_{\ell_2}(v)+E_{\ell_3}(v)=0,\ \ \ \ell_1,\ell_2,\ell_3\ \text{form the boundary of a triangle }f\nonumber\\
&&E_{\ell_1}(v)\cdot E_{\ell_2}(v)=E_{\ell_1}(v')\cdot E_{\ell_2}(v')\ \ \ \ \forall\ell_1,\ell_2\subset t_e\ \text{shared by $\sig_v$ and $\sig_{v'}$.}
\ee
$\eps=\pm1$ is a global sign on the entire triangulation.

\end{Theorem}

{\startproof The above discussion have shown that a critical configuration $(j_f,g_{ve},z_{vf})$ satisfying Eq.\eqref{proddet} implies a nondegenerate $(\pm_vE_\ell(v),\eps)$. So we only need to show that a set of data $(\pm_vE_\ell(v),\eps)$ with $V_4(v)\neq0$ implies a gauge equivalence class of the critical data $(j_f,g_{ve},z_{vf})$ satisfying Eq.\eqref{proddet}.

Given a set of data $(\pm_vE_\ell(v),\eps)$ (with the sign ambiguity $\pm_v$ at each $v$), where $E_\ell(v)$ is a nondegenerate discrete cotetrad on the triangulation, we firstly define the geometrical bivector $X_f(v)$ by $X_f(v):=\eps*\lt[E_{\ell_1}(v)\wedge E_{\ell_2}(v)\rt]$. Thus $\g j_f$ is defined by $\half |X_f(v)|$ which is independent of the choice of $v$ by Eq.\eqref{EOE}.

A spin connection $\O_{vv'}\in SO(1,3)$ is determined uniquely by $E_\ell(v)$ (a proof can be found in \cite{deficit} see also \cite{HZ,CF}) with
\be
E_\ell(v)= \O_{vv'}E_\ell(v'),\ \ \ \forall\ell\subset t_e.\label{EOE}
\ee

A discrete tetrad $U_e(v)$, which corresponds to 5 time-like vectors at each $v$, is given by $E_\ell(v)$ (determined up to a sign ambiguity $\pm_v$ at each $v$) via
\be
U^{e_k}_I(v)=\frac{1}{3!V_4(v)}\sum_{l,m,n}\eps^{jklmn}(v)\eps_{IJKL}E^J_{e_le_j}(v)E^K_{e_me_j}(v)E^L_{e_ne_j}(v)
\ee
where the dependence of $\eps_{jklmn}(v)$ on $v$ reflects the (topological) orientations on different 4-simplices. For example for two neighboring 4-simplices sharing a tetrahedron $t_{e_0}$ determined by $(p_1,p_2,p_3,p_4)$, then $\eps_{01234}(v)=-\eps_{012345}(v')$, see \cite{HZ} for a detailed discussion. Therefore given $e=(v,v')$ dual to a tetrahedron $t_e$
\be
-\mathrm{sgn}(V_4(v)V_4(v'))\frac{U_e(v)}{|U_e(v)|}=\O_{vv'}\frac{U_e(v')}{|U_e(v')|}.
\ee
We again associate a sign $\eps_e(v)=\pm1$ to each vector $U_e(v)$ by requiring $\eps_e(v)=+1$ if $U_e(v)$ is future-directed and $\eps_e(v)=-1$ if $U_e(v)$ is past-directed\footnote{In the asymptotics of the Euclidean EPRL spinfoam amplitude, the sign $\eps_e(v)=\pm1$ has to be an independent variable to characterize the critical data, since the notion of future/past is not an invariant under the action of SO(4). Thus a Euclidean EPRL/FK critical data $(j_f,g_{ve},\hat{n}_{ef})$ is equivalent to the geometrical data $(\pm_vE_\ell(v),\eps,\pm_v\eps_e(v))$.}.

We define $N_e(v)=\eps_e(v)\frac{U_e(v)}{|U_e(v)|}$ (where the $\pm_v$ ambiguities from $\eps_e(v)$ and $U_e(v)$ cancels) which is a set of future-pointing time-like unit 4-vectors. We define a sign $\mu_e=-\eps_e(v)\eps_e(v')\mathrm{sgn}(V_4(v)V_4(v'))$ associated with each edge $e=(v,v')$, and define the spinfoam edge holonomy $g_{vv'}=\mu_e\O_{vv'}$ in the vector representation. One can verify that $N_e(v)$ satisfies $N_e(v)=g_{vv'}N_e(v')$. Thus $g_{vv'}$ preserves the time-direction in the Minkowski space, i.e. $g_{vv'}$ belongs to the restricted Lorentz group $g_{vv'}\in \mathrm{SO}^+(1,3)$.

Given an edge $e$ with $v=s(e), v'=t(e)$, the 4-vector $N_e(v)$ specifies a point $x$ in the
hyperboloid $H$ of future directed unit time-like vectors in Minkowski space. The point $x$ determines a unique pure boost $g_{ve}\in \mathrm{SO}^+(1,3)$, such that $g_{ve}(1,0,0,0)^t=N_e(v)$. Then we define $g_{v'e}=g_{v'v}g_{ve}$ where $g_{v'v}=g_{vv'}^{-1}\in\mathrm{SO}^+(1,3)$. It is easy to verify $g_{v'e}(1,0,0,0)^t=N_e(v')$. We can make a gauge transformation by $g_{ve}\mapsto g_{ve}h_e ,g_{v'e}\mapsto g_{v'e}h_e$, $h_e\in\Su$, which leaves $N_e(v),N_e(v')$ invariant. We identify $g_{ve},g_{v'e}\in\mathrm{SO}^+(1,3)$ to be the vector representation of the critical spinfoam group variable, modulo the SU(2) gauge transformation on the edge. When we lift $g_{ve},g_{v'e}$ to $\Slc$, we obtain the ambiguity because $\Slc$ is the double-cover group to $\mathrm{SO}^+(1,3)$. Again such an ambiguity is a discrete gauge freedom of the spinfoam action $S$.

The bivector $X_f(v)=\eps*\lt[E_{\ell_1}(v)\wedge E_{\ell_2}(v)\rt]$ obviously satisfies the simplicity condition $N_e(v)\cdot *\!X_f(v)=0$ and the parallel transportation $g_{v'v}X_f(v)g_{vv'}=X_f(v')$. If we define $X_{ef}=g_{ev}X_{f}(v)g_{ve}=g_{ev'}X_{f}(v')g_{v'e}$, then $X_{ef}$ satisfies $(1,0,0,0)^t\cdot *\!X_{ef}=0$, which implies there exists a unit 3-vector $\hat{n}_{ef}\in S^2$ such that $*\!X_{ef}=2\g j_f\hat{n}_{ef}\cdot \vec{J}$ if we view $X_{ef}$ belonging to the Lorentz Lie algebra. Similarly we have $*\!X_{e'f}=2\g j_f\hat{n}_{e'f}\cdot \vec{J}$ with $e'$ being the other edge adjacent to $v$. By the definitions of $X_{ef},X_{e'f}$, we have the relation $X_{ef}=g_{ev}g_{ve'}X_{e'f}g_{e'v}g_{ve}$. We define two SO(3) rotations $R_{ef},R_{e'f}$ (determined up to U(1)) by 
\be
R_{ef}\hat{\mathbf{z}}=\hat{n}_{ef},\ \ \ R_{e'f}\hat{\mathbf{z}}=\hat{n}_{e'f}
\ee 
where $\hat{\mathbf{z}}$ is the unit vector along the spatial $z$-direction. Then 
\be
R_{ef}\lt(2\g j_f\hat{\mathbf{z}}\cdot \vec{J}\rt)R_{ef}^{-1}=g_{ev}g_{ve'}R_{e'f}\lt(2\g j_f\hat{\mathbf{z}}\cdot \vec{J}\rt)R_{e'f}^{-1}g_{e'v}g_{ve}
\ee
Thus $R_{ef}^{-1}g_{ev}g_{ve'}R_{e'f}\in\mathrm{SO}^+(1,3)$ is a Lorentz transformation leaving the $xy$-plane invariant, i.e. it is a pure boost in $tz$-plane and a rotation in $xy$-plane, i.e.
\be
g_{ev}g_{ve'}=R_{ef}e^{\b\hat{\mathbf{z}}\cdot \vec{K}+\phi\hat{\mathbf{z}}\cdot \vec{J}}R_{e'f}^{-1},\ \ \ \b,\phi\in\R
\ee
By the complex parametrization of 2-sphere, the unit 3-vectors $\hat{n}_{ef},\hat{n}_{e'f}$ define the normalized spinors $\zeta_{ef},\zeta_{e'f}$ up to a $U(1)$ phase. Then the rotations $R_{ef},R_{e'f}$ and the boost $e^{\b\hat{\mathbf{z}}\cdot \vec{K}}$ are represented on the 2-spinors by the following $\Slc$ matrices
\be
R_{ef}=\left(\begin{array}{cc}\zeta_{ef}^0 & -\bar{\zeta}_{ef}^1 \\ \zeta_{ef}^1 & \bar{\zeta}_{ef}^0\end{array}\right),\ \ 
R_{e'f}=\left(\begin{array}{cc}\zeta_{e'f}^0 & -\bar{\zeta}_{e'f}^1 \\ \zeta_{e'f}^1 & \bar{\zeta}_{e'f}^0\end{array}\right),\ \ 
e^{\b\hat{\mathbf{z}}\cdot \vec{K}+\phi\hat{\mathbf{z}}\cdot \vec{J}}=\left(\begin{array}{cc}e^{\b/2}e^{i\phi/2} & 0\\ 0 & e^{-\b/2}e^{-i\phi/2}\end{array}\right)
\ee
which implies that
\be
g_{ev}g_{ve'}\zeta_{e'f}=e^{\b/2}e^{i\phi/2}\zeta_{ef},\ \ \ g_{ev}g_{ve'}J\zeta_{e'f}=e^{-\b/2}e^{-i\phi/2}J\zeta_{ef}.
\ee
From the second above equation, we obtain that $(g_{ve'}^{\dagger})^{-1}\zeta_{e'f}$ equals $(g_{ve}^{\dagger})^{-1}\zeta_{ef}$ up to a complex rescaling. We thus construct the critical spinfoam variable $z_{vf}$ by the normalization
\be
z_{vf}=e^{-i\phi^f_{e'v}}\frac{(g_{ve'}^{\dagger})^{-1}\zeta_{e'f}}{||(g_{ve'}^{\dagger})^{-1}\zeta_{e'f}||}=e^{-i\phi^f_{ev}}\frac{(g_{ve}^{\dagger})^{-1}\zeta_{e'f}}{||(g_{ve}^{\dagger})^{-1}\zeta_{ef}||}
\ee
with $\phi^f_{e'v}-\phi^f_{ev}=\phi/2$. Therefore we have reconstructed the spinfoam data $(j_f,g_{ve},z_{vf})$ from the geometrical data $(\pm_v E_\ell(v),\eps)$. It is a straight-forward task to check that the reconstructed spinfoam data does satisfy the critical equations.
\finishproof}

The above equivalence theorem allows us to study the properties of the spinfoam critical configurations via its simplicial geometry correspondence.

\begin{Proposition}\label{sols}

A Regge-like spin configuration $j_f$ on a simplicial complex results in a finite number of spinfoam critical configurations $(j_f,g_{ve},z_{vf})$ satisfying the nondegenercy assumption.

\end{Proposition}

{\startproof A spin configuration $j_f$ is called Regge-like if there exists a set of edges-lengths $s_\ell$ on the simplicial complex, such that $\g j_f$ are areas expressed by the edge-lengths. In general a Regge-like spin configuration may result in more than one set of edge-lengths. On a simplicial complex, there are only a finite number $\#_s$ of different set of edge-lengths resulting from a Regge-like areas.

Given a set of edge-lengths $s_\ell$, up to $\Slc$ gauge transformation and global sign $\pm_v$, it determines two cotetrads, $E_\ell(v)$ and its parity inversion $\tilde{E}_\ell(v)=\mathbf{P}E_\ell(v)$, at each vertex $v$, as it was discussed in the last section. Therefore a set of edge-lengths $s_\ell$ results in $2^{\#_v}$ cotetrads on the simplicial complex. Including the global sign ambiguity $\eps$, less than $2\times2^{\#_v}$ possible geometrical configurations $(\pm_vE_\ell(v),\eps)$ are determined by a set of edge-lengths. The corresponding spinfoam critical data $(j_f,g_{ve},z_{vf})$ is described by the solution-generating maps defined in Section \ref{SG}.

Therefore given a Regge-like spin configuration $j_f$, up to gauge transformation, by Theorem \ref{equivalent}, the number of critical configurations $(j_f,g_{ve},z_{vf})$ satisfying the nondegenercy assumption is given by $2\times2^{\#_v}\times\#_s$.
\finishproof}

There are two remarks:

\begin{itemize}

\item Note that the local parity inversion $\mathbf{P}$, which transforms a critical configuration to another, flips the local spacetime orientations $\mathrm{sgn}(V_4(v))$ of the individual 4-simplices. Thus there exists many spinfoam critical configurations corresponding to the non-uniform spacetime orientation $\mathrm{sgn}(V_4(v))$. Given a set of edge-length implied by $j_f$, there are only 2 critical configurations with uniform spacetime orientation, i.e. one with $\mathrm{sgn}(V_4(v))>0$ everywhere and the other with $\mathrm{sgn}(V_4(v))<0$ everywhere.

\item If the simplicial complex has a boundary, it is shown in \cite{HZ} that the global sign $\eps$ is determined by specifying the 3-dimensional boundary orientation. Therefore in this case only a half of the number of critical configurations presents. 

\end{itemize}

The above discussion respects the nondegeneracy assumption Eq.\eqref{proddet}. We have developed the classification and counting of the spinfoam critical configurations, which corresponds to nondegenerate Lorentzian simplicial geometry. For the critical configurations violating Eq.\eqref{proddet}, the analysis is done by an analog of the above discussion, combining the corresponding discussion in \cite{HZ}. Such an analysis is not the main purpose of this paper and is presented in Appendix \ref{DEG}.

\section{Solution-Generating Maps}\label{SG}

\noindent {\underline{Local Parity Inversion}}:

As it was mentioned previously, there is a parity transformation $\Fp$ between spinfoam critical configurations with $\Fp^2=\id$
\be
\Fp: (j_f,g_{ve},z_{vf})\mapsto (j_f,\tilde{g}_{ve},\tilde{z}_{vf})\ \ \ 
\text{where}\ \ \ \tilde{g}_{ve}=Jg_{ve}J^{-1}=(g_{ve}^\dagger)^{-1},\ \ \ \tilde{z}_{vf}=\frac{g_{ve}g_{ve}^\dagger z_{vf}}{||g_{ve}^\dagger z_{vf}||^2}
\ee
where the data $(j_f,\tilde{g}_{ve},\tilde{z}_{vf})$ satisfies the critical equations Eqs.\eqref{1}, \eqref{2}, and \eqref{3}. The definition of $\tilde{z}_{vf}$ is independent of the choice of $e$ because of $z_{vf}$ is a critical variable, thus satisfies Eq.\eqref{2}. Such a parity transformation was firstly pointed out by Barrett et al in \cite{semiclassical} in the asymptotic analysis of the spinfoam vertex amplitude. An important property of the parity transformation is that the transformation is \emph{local}, because it leaves $g_{ve}^\dagger z_{vf}/||g_{ve}^\dagger z_{vf}||$ invariant (thus the auxiliary spinor $\zeta_{ef}$ is invariant). Therefore one may transform the variables $g_{ve},z_{vf}\mapsto \tilde{g}_{ve},\tilde{z}_{vf}$ at the vertex $v$ but leave the other variables $g_{v'e},z_{v'f}$ at other vertices invariant, while Eq.\eqref{1} still holds. We have the relation:
\be
||\tilde{g}_{ve}^\dagger \tilde{z}_{vf}||=||g_{ve}^\dagger z_{vf}||^{-1}
\ee
Thus $\Fp$ flips between Eqs.\eqref{zeta1} and \eqref{zeta2}. 

Given that the parity transformation $\Fp$ is local, we only need to consider a single 4-simplex in order to see its geometrical implication. We first consider the unit time-like vector $N_e(v)=g_{ve}\rhd(1,0,0,0)^t$. The relation between $N_e(v)$ and its parity transformation $\tilde{N}_e(v)=\tilde{g}_{ve}\rhd(1,0,0,0)^t$ can be shown by using the Hermitian matrix representation of the vectors $V=\frac{1}{\sqrt{2}}[V^0\mathbf{1}+V^j\sig_j]$, thus
\be
\tilde{N}_e(v)=\tilde{g}_{ve}\tilde{g}_{ve}^\dagger=Jg_{vf}g_{vf}^\dagger J^{-1}=JN_e(v)J^{-1}=\frac{1}{\sqrt{2}}\lt[N_e^0(v)\mathbf{1}-N^j_e(v)\sig_j\rt]
\ee
since $J\vec{\sig} J^{-1}=-\vec{\sig}$. We denote the parity inversion in $(\mathbb{R}^4,\eta_{IJ})$ by $\bp=\mathrm{diag}(1,-1,-1,-1)$, then we have $\tilde{N}_e(v)=\bp N_e(v)$.

We define the bivectors $\tilde{X}_f(v)=\tilde{g}_{ve}\otimes\tilde{g}_{ve}\rhd \tilde{X}_{ef}$ within the 4-simplices $\sig_v$, where
\be
\tilde{X}_{ef}^{IJ}=X_{ef}^{IJ}=2\g j_f\lt[u\wedge\hat{n}_{ef}\rt]\ \ \ \ \ u=(1,0,0,0)^t
\ee
since $\zeta_{ef}$ is invariant under $\Fp$. Consider the Hermitian matrix representation of $\hat{n}_{ef}$, the action $\tilde{g}_{ve}\rhd\hat{n}_{ef}$ is given by (note that $J^2=-1$)
\be
\frac{1}{\sqrt{2}}\tilde{g}_{ve}(\hat{n}_{ef}\cdot\vec{\sig})\tilde{g}_{ve}^\dagger=\frac{1}{\sqrt{2}}J{g}_{ve}J^{-1}(\hat{n}_{ef}\cdot\vec{\sig})J{g}_{ve}^\dagger J^{-1}=-\frac{1}{\sqrt{2}}J{g}_{ve}(\hat{n}_{ef}\cdot\vec{\sig}){g}_{ve}^\dagger J^{-1}=-\frac{1}{\sqrt{2}}\bp{g}_{ve}(\hat{n}_{ef}\cdot\vec{\sig}){g}_{ve}^\dagger
\ee
while we have shown $\tilde{g}_{ve}\rhd u=\bp\lt({g}_{ve}\rhd u\rt)$, thus we obtain that 
\be
\tilde{X}_f(v)=-(\bp\otimes\bp){X}_f(v)
\ee
Recall the previous construction towards the geometrical interpretation of critical data. Following the same argument towards Eq.\eqref{NN}, we obtain the following relation between the bivectors and $\tilde{N}_e(v)$'s constructed from $\tilde{g}_{ve}$
\be
\tilde{X}_f(v)=\tilde{\a}_{ee'}(v)\tilde{N}_e(v)\wedge \tilde{N}_{e'}(v)\ \ \Rightarrow\ \ -(\bp\otimes\bp){X}_f(v)=\tilde{\a}_{ee'}(v)\bp{N}_e(v)\wedge \bp{N}_{e'}(v)
\ee
Then we have the relation $\tilde{\a}_{ee'}(v)=-\a_{ee'}(v)$ and $\tilde{\b}_{ee'}(v)=-\b_{ee'}(v)$, where $\b_{ee'}(v)=\a_{ee'}(v)\eps_{ee'}(v)$ and $\tilde{\b}_{ee'}(v)=\tilde{\a}_{ee'}(v)\eps_{ee'}(v)$. Following the same procedure as before, we denote $\tilde{\b}_{e_ie_j}$ by $\tilde{\b}_{ij}$ and construct the closure condition for the 4-simplex $\tilde{\sig}_v$
\be
\sum_{j=1}^5\tilde{\b}_{ij}(v)\tilde{N}_{e_j}(v)=0
\ee
by choosing the nonvanishing diagonal elements $\tilde{\b}_{ii}$. Since we have the closure condition $\sum_{j=1}^5{\b}_{ij}{N}_{e_j}(v)=0$, the parity inversion $\tilde{N}_e(v)=\bp N_e(v)$, and $\tilde{\b}_{ij}(v)=-\b_{ij}(v)$ for $i\neq j$, we obtain that the diagonal elements $\tilde{\b}_{ii}(v)=-{\b}_{ii}(v)$. Furthermore we can show that $\tilde{\b}_{ij}$ can be factorized in the same way as in Lemma \ref{factorization}
\be
\tilde{\b}_{ij}(v)=\mathrm{sgn}(\tilde{\b}_{j_0j_0}(v))\tilde{\b}_i(v)\tilde{\b}_j(v)\ \ \ \ \ \ \tilde{\b}_j(v)={\tilde{\b}_{jj_0}(v)}\big/{\sqrt{|\tilde{\b}_{j_0j_0}(v)|}}
\ee
which results in that
\be
\mathrm{sgn}(\tilde{\b}_{j_0j_0}(v))=-\mathrm{sgn}({\b}_{j_0j_0}(v))\ \ \ \ \text{and}\ \ \ \ \tilde{\b}_j(v)=-{\b}_j(v)
\ee
We construct the 4-volume for $\tilde{\b}_j(v)\tilde{N}_{e_j}(v)$
\be
\tilde{V}_4(v):=\det\Big(\tilde{\b}_2(v)\tilde{N}^{e_2}(v),\tilde{\b}_3(v)\tilde{N}^{e_3}(v),\tilde{\b}_4(v)\tilde{N}^{e_4}(v),\tilde{\b}_5(v)\tilde{N}^{e_5}(v)\Big)=-V_4(v)		
\ee
since the Minkowski parity inversion flips the sign of oriented 4-volume. Recall the sign factor $\eps(v)=\mathrm{sgn}({\b}_{j_0j_0}(v))\mathrm{sgn}(V_4(v))$, then we have for the parity inversion
\be
\tilde{\eps}(v)=\mathrm{sgn}(\tilde{\b}_{j_0j_0}(v))\mathrm{sgn}(\tilde{V}_4(v))=\eps(v)
\ee
Recall that the sign $\eps(v)$ is actually independent of $v$ by Lemma \ref{globalsign}. This result shows that the critical data $(j_f, {g}_{ve},{z}_{vf})$ and its parity transformation $(j_f, \tilde{g}_{ve},\tilde{z}_{vf})$ corresponds to an identical global sign factor $\eps$ for the bivector.

The discrete tetrad is defined by $U_e(v)=\pm_v\frac{\b_e(v)N_e(v)}{\sqrt{|V_4(v)|}}$. $\tilde{U}_e(v)$ constructed from parity configuration $(j_f, \tilde{g}_{ve},\tilde{z}_{vf})$ relates $U_e(v)$ from $(j_f, {g}_{ve},{z}_{vf})$ by a Minkowski parity inversion
\be
\tilde{U}_e(v)=\bp{U}_e(v)
\ee
The same relation holds for the cotetrad $\tilde{E}_{\ell}(v)$
\be
\tilde{E}_{\ell}(v)=\bp{E}_{\ell}(v)
\ee
from the relation $\tilde{U}^{e_j}_I(v)\tilde{E}^I_{e_ke_l}(v)=\delta^j_k-\delta^j_l$. For the geometrical interpretation of bivector
\be
\tilde{X}_f(v)=\eps\ \tilde{V}_4\lt[\tilde{U}_e(v)\wedge \tilde{U}_{e'}(v)\rt]\ \ \ \ \text{and}\ \ \ \ \tilde{X}_f(v)=\eps\ *\lt[\tilde{E}_{\ell_1}(v)\wedge\tilde{E}_{\ell_2}(v)\rt]
\ee
which is consistent because of the relations $\tilde{X}_f(v)=-(\bp\otimes\bp){X}_f(v)$, $\tilde{U}_e(v)=\bp{U}_e(v)$, $\tilde{E}_{\ell}(v)=\bp{E}_{\ell}(v)$, $\tilde{V}_4(v)=-V_4(v)$, and $\eps_{IJKL}\bp^I_{\ M}\bp^J_{\ N}\bp^K_{\ P}\bp^L_{\ Q}=-\eps_{MNPQ}$. Here we emphasize that the sign factor $\eps$ for the parity configuration $(j_f, \tilde{g}_{ve},\tilde{z}_{vf})$ is the same as the original configuration $(j_f, g_{ve},{z}_{vf})$, thus is consistent with the fact that $\eps$ is a global sign factor on the entire simplicial complex, i.e. the local parity inversion of the critical data doesn't change the sign $\eps$ locally.

Now we summarize the geometrical implication of the local parity inversion $\Fp$. Given a critical data $(j_f, g_{ve},{z}_{vf})$ which is equivalent to the nondegenerate geometrical data $(\pm_vE_\ell(v);\eps)$, a parity inversion $\Fp$ acting locally on the variables at a single vertex $v$ (or several vertices) gives a new critical data $(j_f, \tilde{g}_{ve},\tilde{z}_{vf})$ where $\tilde{g}_{ve}\neq g_{ve},\tilde{z}_{vf}\neq z_{vf}$ only at $v$, where the new critical data is equivalent to the geometrical data $(\pm_v\bp E_\ell(v),\{\pm_{v'}E_\ell(v')\}_{v'\neq v};\eps)$ with $\bp$ the parity inversion on Minkowski space.

\vspace{0.5cm}

\noindent {\underline{Global $J$-Parity Inversion}}:

We define another solution-generating map $\Fj$ between critical configurations with $\Fj^2=1$ by the following
\be
\Fj:(j_f,g_{ve},z_{vf})\mapsto (j_f,\check{g}_{ve},\check{z}_{vf})\ \ \text{where}\ \ \check{g}_{ve}=Jg_{ve}J^{-1}=(g_{ve}^\dagger)^{-1},\ \ \check{z}_{vf}=iJz_{vf}
\ee 
for \emph{all} the $g_{ve},z_{vf}$ variables, which we call a \emph{global} $J$-Parity inversion. Such a transformation doesn't leaves $g^\dagger_{ve}z_{vf}/||g^\dagger_{ve}z_{vf}||$ invariant thus transforms the auxiliary spinor $\zeta_{ef}$ into
\be
\check{\zeta}_{ef}=e^{i\check{\phi}^f_{ev}}\frac{\check{g}^\dagger_{ve}\check{z}_{vf}}{\lt|\lt|\check{g}^\dagger_{ve}\check{z}_{vf}\rt|\rt|}=iJ \lt[e^{-i\check{\phi}^f_{ev}}\frac{{g}^\dagger_{ve}{z}_{vf}}{\lt|\lt|{g}^\dagger_{ve}{z}_{vf}\rt|\rt|}\rt]=iJ \zeta_{ef}
\ee
if we ask $\check{\phi}^f_{ev}=-{\phi}^f_{ev}$. Here we also use the relation that $\lag JZ,JZ'\rag=\lag Z',Z\rag$, i.e. $J$ is anti-unitary. It is straight-forward to check the critical equations Eqs.\eqref{1}, \eqref{2} and \eqref{3} are invariant if we change $\a_{vv'}\mapsto\check{\a}^f_{vv'}=-\a^f_{vv'}$. It means that given a solution $(j_f,g_{ve},z_{vf})$ of the critical equation with the phase $e^{i\a^f_{vv'}}$ in Eq.\eqref{1}, the map $\Fj$ gives a new solution $(j_f,\check{g}_{ve},\check{z}_{vf})$ of the critical equation with the phase $e^{-i\a^f_{vv'}}$ in Eq.\eqref{1}. Moreover such a transformation must act globally on all the $g_{ve},z_{vf}$ variables in order to make Eq.\eqref{1} invariant.

Since the expression of $\check{g}_{ve}$ is the same as $\tilde{g}_{ve}$ in the above parity transformation $\Fp$, the unit time-like vector $N_e(v)$ transforms in the same way as above, i.e. 
\be
N_{e}(v)\mapsto \check{N}_e(v)=\bp N_e(v)
\ee
where $\bp$ is the parity inversion on Minkowski spacetime. However the bivectors $X_{ef}$ don't transform in the same way as above, because $\hat{n}_{ef}\mapsto -\hat{n}_{ef}$ since $\zeta_{ef}\mapsto iJ\zeta_{ef}$ by the action of $\Fj$. So here we have
\be
X_{ef}\mapsto\check{X}_f(v)=(\bp\otimes\bp){X}_f(v)
\ee
which implies $\check{\b}_{ee'}=\b_{ee'}$ thus $\mathrm{sgn}(\check{\b}_{j_0j_0})=\mathrm{sgn}({\b}_{j_0j_0})$. On the other hand, since $\check{N}_e(v)=\bp N_e(v)$, the oriented 4-volume flips its sign $\check{V}_4(v)=-{V}_4(v)$. Therefore we obtain that the global sign $\eps$ flips by the action of $\Fj$, i.e.
\be
\eps\mapsto \check{\eps}=-\eps
\ee
which is consistent with the fact that the $J$-parity inversion $\Fj$ is defined globally on all the $g_{ve},z_{vf}$ variables. 

By the same construction as above, we obtain the tetrad and cotetrad transformed by $J$-parity inversion $\Fj$ is the same as the ones transformed by the parity inversion $\Fp$, i.e.
\be
\check{U}_e(v)=\bp U_e(v),\ \ \ \check{E}_\ell(v)=\bp E_\ell(v)
\ee
Then the bivector is expressed as 
\be
\check{X}_f(v)&=&\check{\eps}\ \check{V}_4\lt[\check{U}_e(v)\wedge \check{U}_{e'}(v)\rt]={\eps}\ {V}_4\lt[\bp{U}_e(v)\wedge \bp{U}_{e'}(v)\rt]\nonumber\\
\check{X}_f(v)&=&\check{\eps}\  *\lt[\check{E}_{\ell_1}(v)\wedge\check{E}_{\ell_2}(v)\rt]=-{\eps}\  *\lt[\bp{E}_{\ell_1}(v)\wedge\bp{E}_{\ell_2}(v)\rt].
\ee

As a result, we summarize the geometrical implication of the global $J$-parity inversion $\Fj$. Given a critical data $(j_f, g_{ve},{z}_{vf})$ which is equivalent to the nondegenerate geometrical data $(\pm_vE_\ell(v);\eps)$, a global $J$-parity inversion $\Fj$ acting globally on all the $g_{ve},{z}_{vf}$ variables gives a new critical data $(j_f, \check{g}_{ve},\check{z}_{vf})$, where the new critical data is equivalent to the geometrical data $(\pm_v\bp E_\ell(v);-\eps)$ with $\bp$ the parity inversion on Minkowski space.

\vspace{0.5cm}

As it was mentioned in the proof of Proposition \ref{sols}, given a set of edge-lengths implied by the Regge-like $j_f$, there are in total $2^{\#_v}\times2$ spinfoam critical configurations satisfying the nondegeneracy conditions Eq.\eqref{proddet}. It is shown above that given any one of these $2^{\#_v}\times2$ spinfoam critical configurations, all the others can be completely generated by the solution-generating maps: local parity inversion $\Fp$ and global $J$-parity inversion $\Fj$. It is shown in the following commutative diagram:
\be
&\text{\small Local Parity Inversion $\Fp$}&\nonumber\\
(j_f,g_{ve},z_{vf})\Leftrightarrow(\pm_vE_\ell(v);\eps)&{\xrightarrow{\hspace*{5cm}}}&(j_f,\tilde{g}_{ve},\tilde{z}_{vf})\Leftrightarrow(\pm_v\bp E_\ell(v);\eps)\nonumber\\
\downarrow\ \ \ \ \ \ \ \ \ \ \ \ \ \ \ \ \ &&\ \ \ \ \ \ \ \ \ \ \ \ \ \ \ \ \downarrow\nonumber\\
\text{\small Global $J$-Parity Inversion $\Fj$}\ \ \ &&\ \ \ \text{\small Global $J$-Parity Inversion $\Fj$}\nonumber\\
\downarrow\ \ \ \ \ \ \ \ \ \ \ \ \ \ \ \ \ &&\ \ \ \ \ \ \ \ \ \ \ \ \ \ \ \ \downarrow\nonumber\\
(j_f,\check{g}_{ve},\check{z}_{vf})\Leftrightarrow(\pm_v\bp E_\ell(v);-\eps)&{\xrightarrow{\hspace*{5cm}}}&(j_f,\check{\tilde{g}}_{ve},\check{\tilde{z}}_{vf})\Leftrightarrow(\pm_vE_\ell(v);-\eps)\nonumber\\
&\text{\small Local Parity Inversion $\Fp$}&\nonumber\\
\ee
where the local parity $\Fp$ and global $J$-parity $\Fj$ indeed commute to each other since $\check{\tilde{g}}_{ve}=\tilde{\check{g}}_{ve}=g_{ve}$ and
\be
\tilde{\check{z}}_{vf}=\frac{\check{g}_{ve}\check{g}_{ve}^\dagger\check{z}_{vf}}{||\check{g}_{ve}^\dagger\check{z}_{vf}||^2}=iJ\lt[\frac{{g}_{ve}{g}_{ve}^\dagger{z}_{vf}}{||{g}_{ve}^\dagger{z}_{vf}||^2}\rt]=iJ\tilde{z}_{vf}=\check{\tilde{z}}_{vf}.
\ee

We remark that the global $J$-parity inversion, as a symmetry of the critical equations, is broken explicitly by the boundary of the spinfoam amplitude, when we impose an orientation of the boundary, because it is shown in \cite{HZ} that the boundary orientation fix the sign ambiguity $\eps$\footnote{It can also be seen by the fact that $\zeta_{ef}$ is the boundary data if there is a boundary of spinfoam, while the global $J$-parity inversion transform $\zeta_{ef}\mapsto iJ\zeta_{ef}$.}.

\section{Time-Oriented Critical Configurations}\label{timeorientation}

The equivalent relation between spinfoam critical configurations and simplicial geometries described in Theorem \ref{equivalent} allows us to study the spacetime time-orientation in the context of large spin spinfoam asymptotics. 

In standard GR, given a 4-dimensional spacetime with a spacetime-orientation $(\cm,g_{\a\b},\epsilon_{\a\b\mu\nu})$, its oriented orthonormal frame bundle is a principal fiber bundle with structure group $\mathrm{SO}(1,3)$ where the orthonormal basis satisfies $\epsilon_{\a\b\mu\nu}e_0^\a e^\b_1 e_2^\mu e_3^\nu>0$. Given a 4-dimensional spacetime with both a spacetime-orientation and a time-orientation $(\cm,g_{\a\b},\epsilon_{\a\b\mu\nu},T)$ where $T$ is a time-function, its oriented and time-oriented orthonormal frame bundle is a principle fiber bundle with structure group $\mathrm{SO}^+(1,3)$ or $\Slc$, since the orthonormal basis has an additional constraint $e^\a_0\nabla_\a T<0$. Then in order to preserve the spacetime-orientation and time-orientation of the orthonormal basis, the holonomies of any path is a group element in $\Slc$. 

For a spinfoam critical data $(j_f,g_{ve},z_{vf})$ equivalent to the geometrical data $(\pm_v E_\ell(v),\eps)$, its spacetime-orientation is specified by $\mathrm{sgn}(V_4(v))$. Since we have the parallel transportation $E_\ell(v)=\O_{vv'}E_\ell(v)$ with the spin connection $\O_{vv'}=\mu_e g_{vv'}$ ($\mu_e=-\eps_e(v)\eps_e(v')\mathrm{sgn}(V_4(v)V_4(v'))$), then as an analog of the standard GR case, it seems we should require the spin connection $\O_{vv'}$ belongs to $\mathrm{SO}^+(1,3)$ to define a discrete analog of time-oriented spacetime. However such a requirement doesn't make sense in our context because the discrete cotetrad $E_\ell(v)$ is only determined up to the sign ambiguity $\pm_v$. Thus $\O_{vv'}$ also has a sign ambiguity which ruins the requirement that $\O_{vv'}\in\mathrm{SO}^+(1,3)$. Therefore we have to look for the quantity which is free of the $\pm_v$-ambiguity. A good candidate is the loop holonomy of spin connection $\O_f=\vec{\prod}_{e\subset\partial f}\O_e$ along the boundary of a dual face. So we make the following definition:

\begin{Definition}

Considering a spinfoam critical data $(j_f,g_{ve},z_{vf})$ equivalent to the nondegenerate geometrical data $(\pm_v E_\ell(v),\eps)$:

(1) Given a region $\calr$ in the simplicial complex with a constant spacetime orientation $\mathrm{sgn}(V_4(v))$, the region $\calr$ is time-oriented if for all the faces $f$ in the region, the loop holonomy of spin connection $\O_f$ belongs to $\mathrm{SO}^+(1,3)$.

(2) Since the spinfoam variable $g_{ve}$ always belongs to $\Slc$, then an equivalent definition is that for all the faces $f$ in the region $\calr$\footnote{The definition Eq.\eqref{epseps} may be understood as a generalization of condition $e^\a_0\nabla T<0$ to our context, once we consistently choose a future-direction in the internal Minkowski space where $U_e(v)$ lives.}
\be
\prod_{e\subset\partial f}\mu_e=\prod_{e\subset\partial f}\lt[-\eps_e(v)\eps_e(v')\rt]=1\label{epseps}
\ee
Recall that $\eps_e(v)=+1$ or $-1$ if the discrete tetrad $U_e(v)$ is future-pointing or past-pointing in the Minkowski space.

\end{Definition}

In order to construct the oriented and time-oriented spacetime region as a spinfoam critical configuration, we define a nondegenerate discrete time-like tetrad $U_e(v)$ at each vertex satisfying $\sum_e U_e(v)=0$ (specifying a tetrad $U_e(v)$ is equivalent to specifying a cotetrad $E_\ell(v)$), and we ask 

\begin{enumerate}
	\item For each vertex $v$ in a region $\calr$, the (inverse) 4-volume $\frac{1}{V_4(v)}=\det\lt(U^{e_2}(v),U^{e_3}(v),U^{e_4}(v),U^{e_5}(v)\rt)$ is uniformly positive (or uniformly negative).
	
	\item For each edge $e$ connecting $v,v'$ in $\calr$, the two time-like vectors $U_e(v)$ and $U_e(v')$ contain one future-pointing vector and one past-pointing vector.
\end{enumerate}
\noindent
The above second requirement immediately implies that $\lt[-\eps_e(v)\eps_e(v')\rt]=1$ thus Eq.\eqref{epseps} is satisfied for the tetrad $U_e(v)$ on the simplicial subcomplex $\calr$. Eq.\eqref{epseps} is certainly not going to be violated if we implement the ambiguity $\pm_v$ to $U_e(v)$ at each individual vertex. If $E_\ell(v)$ is the cotetrad uniquely determined by the tetrad $U_e(v)$ satisfying the above requirements (up to $\pm_v$), then by the equivalence theorem \ref{equivalent}. The spinfoam critical data $(j_f,g_{ve},z_{vf})$ equivalent to $(\pm_v E_\ell(v),\eps)$ is a spinfoam analog of the oriented and time-oriented spacetime in the region $\calr$.

\section{Critical Action}\label{CriticalAction}

The spinfoam action $S$ evaluated at the critical configuration $(j_f,g_{ve},z_{vf})$, written on the exponential $e^{S(x_0)}$, is the overall factor in the asymptotic expansion given in Theorem \ref{asymptotics}, and gives the leading contribution to the asymptotic series. The asymptotic expansion develops a perturbation theory of the partial amplitude $A_{j_f}(\ck)$ from the vacuum $(j_f,g_{ve},z_{vf})$. Here we consider the critical action at a physical vacuum, which is a spinfoam critical configuration corresponding to an oriented, time-oriented, non-degenerate Lorentzian simplical geometry.

The spinfoam action $S$ can be written as a sum of face actions $S_f$, which are evaluated at the critical configuration solving Eqs.\eqref{1}, \eqref{2}, and \eqref{3} or Eqs.\eqref{zeta1}, \eqref{zeta2}, and \eqref{zeta3}:
\be
S_f=2i\g j_f\sum_{v\in\partial f}\ln\frac{||Z_{ve'f}||}{||Z_{vef}||}-2i j_f\sum_{v\in\partial f}\phi_{eve'}=-2i j_f\lt(\g\sum_{v\in\partial f}\theta_{eve'} +\sum_{v\in\partial f}\phi_{eve'}\rt)
\ee
where we have denoted ${||Z_{vef}||}/{||Z_{ve'f}||}:=e^{\theta_{eve'}}$. Recall Eqs.\eqref{zeta1} and \eqref{zeta2}, and consider the following successive actions on $(\zeta_{ef},J\zeta_{ef})$ of $g_{e'v}g_{ve}$ along the entire boundary of the face $f$
\be
\overleftarrow{\prod_{v\in\partial f}}g_{e'v}g_{ve}\lt(\zeta_{ef},J\zeta_{ef}\rt)&=&\lt(e^{\sum_{v}\theta_{eve'}+i\sum_v\phi_{eve'}}\zeta_{ef},e^{-\sum_{v}\theta_{eve'}-i\sum_v\phi_{eve'}}J\zeta_{ef}\rt)
\ee
So if we consider $(\zeta_{ef},J\zeta_{ef})$ is an SU(2) matrix, we obtain the expression of the loop holonomy $G_f(e)=\overleftarrow{\prod}_{v\in\partial f}g_{e'v}g_{ve}$
\be
G_f(e)=\exp\lt[\sum_{v\in\partial f}\lt(\theta_{eve'}+i\phi_{eve'}\rt)\vec{\sig}\cdot \hat{n}_{ef}\rt].
\ee
where $\hat{n}_{ef}=\langle \zeta_{ef}|\vec{\sig}|\zeta_{ef}\rangle$.
%which is an exponential map from Lie algebra variable\footnote{Note that not all the elements in $\Slc$ can be written in an exponential form, because of the noncompactness.}.

We have the following identities
\be
\ln\tr\lt[\half\lt(1+\vec{\sig}\cdot\hat{n}_{ef}\rt)G_f(e)\rt]&=&\sum_{v\in\partial f}\theta_{eve'}+i\sum_{v\in\partial f}\phi_{eve'}\nonumber\\
\ln\tr\lt[\half\lt(1+\vec{\sig}\cdot\hat{n}_{ef}\rt)G_f^\dagger(e)\rt]&=&\sum_{v\in\partial f}\theta_{eve'}-i\sum_{v\in\partial f}\phi_{eve'}
\ee
by using the properties of Pauli matrices: $(\vec{\sig}\cdot\hat{n})^{2k}=1_{2\times 2}$ and $(\vec{\sig}\cdot\hat{n})^{2k+1}=\vec{\sig}\cdot\hat{n}$. Insert these relations into the expression of the face action $S_f$
\be
S_f%&=&-2i j_f\g\sum_{v}\theta_{eve'} -2i j_f\sum_v\phi_{eve'}\nonumber\\
%&=&-i j_f\g\lt\{\ln\tr\lt[\half\lt(1+\vec{\sig}\cdot\hat{n}_{ef}\rt)G_f(e)\rt]+\ln\tr\lt[\half\lt(1+\vec{\sig}\cdot\hat{n}_{ef}\rt)G^\dagger_f(e)\rt]\rt\} \nonumber\\
%&&\ -j_f\lt\{\ln\tr\lt[\half\lt(1+\vec{\sig}\cdot\hat{n}_{ef}\rt)G_f(e)\rt]-\ln\tr\lt[\half\lt(1+\vec{\sig}\cdot\hat{n}_{ef}\rt)G^\dagger_f(e)\rt]\rt\}\nonumber\\
&=&-(i\g+1)j_f\ln\tr\lt[\half\lt(1+\vec{\sig}\cdot\hat{n}_{ef}\rt)G_f(e)\rt]-(i\g-1)j_f\ln\tr\lt[\half\lt(1+\vec{\sig}\cdot\hat{n}_{ef}\rt)G^\dagger_f(e)\rt]
\ee
We define the following variables by making a parallel transport to a vertex $v$
\be
\hat{X}_f(v):=g_{ve}\vec{\sig}\cdot\hat{n}_{ef}g_{ev}&\ \ \ \ &\hat{X}^\dagger_f(v):=g_{ev}^\dagger\vec{\sig}\cdot\hat{n}_{ef}g_{ve}^\dagger\nonumber\\
G_f(v):=g_{ve} G_f(e) g_{ev}\ &\ \ \ \ &G_f^\dagger(v):=g_{ev}^\dagger G_f(e) g_{ve}^\dagger
\ee
From Eq.\eqref{Xsig} $\hat{X}_f(v)$ is related to the bivector $X_f(v)$ by $\hat{X}_f(v)={X}_f(v)/\g j_f$. Clearly the definition of $\hat{X}_f(v)$ is independent of the choice of $e$ by the discussion in Section \ref{bivector}. In terms of $\hat{X}_f(v)$ and $G_f(v)$, the critical face action is written as
\be
S_f=-(i\g+1)j_f\ln\tr\lt[\half\lt(1+\hat{X}_f(v)\rt)G_f(e)\rt]-(i\g-1)j_f\ln\tr\lt[\half\lt(1+\hat{X}_f^\dagger(v)\rt)G^\dagger_f(e)\rt]\label{Sf0}
\ee
According to the geometrical interpretation of the critical data, there exists a unique cotetrad $E_\ell(v)$ up to $\pm_v$, such that the bivector $\hat{X}_f(v)$ is written as
\be
\hat{X}_f(v)=2\eps\frac{*E_{\ell_1}(v)\wedge E_{\ell_2}(v)}{|*E_{\ell_1}(v)\wedge E_{\ell_2}(v)|}
\ee
and the spinfoam edge holonomy $g_{vv'}$ equals to the spin connection $\O_{vv'}$ up to a sign $\mu_e$, i.e. $g_{vv'}=\mu_e \O_{vv'}$. We assume that the critical data corresponds to a time-oriented spacetime, then $\prod_{e\subset \partial f}\mu_e=1$. Thus we have 
\be
G_f(v)=\O_f(v)=\overleftarrow{\prod_{(v,v')\subset\partial f}}\O_{vv'}.
\ee
Therefore the spinfoam loop holonomy at the critical point satisfies
\be
G_f(v)E_{\ell}(v)=E_{\ell}(v)\label{GE}
\ee
for the three edges $\ell$ of the triangle $f$. The spinfoam loop holonomy $G_f(v)\in\mathrm{SO}^+(1,3)$ then can only be a pure boost leaving the triangle $f$ invariant. Then $G_f(v)$ can be expressed as 
\be
G_f(v)=\exp\lt[\frac{* E_{\ell_1}(v)\wedge E_{\ell_2}(v)}{|*E_{\ell_1}(v)\wedge E_{\ell_2}(v)|}\Theta_f\rt]
=e^{\eps\frac{1}{2}\Theta_f\hat{X}_f(v)}
\ee
where $\Theta_f$ is the Regge deficit angle\footnote{$\Theta_f$ being the Regge deficit angle relies on the assumption that the critical data corresponds to a globally oriented discrete spacetime, i.e. $\mathrm{sgn}(V_4(v))$ is a constant in the region we are interested in. See \cite{HZ} for a detailed discussion.} determined by the geometry $E_\ell(v)$ \cite{deficit}

Inserting the expression of $G_f(v)$ into Eq.(\ref{Sf0}), we obtain the expression of critical action\footnote{In case that there is a boundary of the spinfoam, the critical action at a physical vacuum has an additional Regge boundary term, i.e. $S=-i\eps \sum_f\g j_f\Theta_f-i\eps \sum_f\g j_f\Theta^B_f$, where $\Theta^B_f$ is the 4-d dihedral angle between the corresponding boundary tetrahedra. Here the global sign $\eps$ is either $+1$ or $-1$ is fixed by the 3-d orientation from boundary data. }
\be
S=-i\eps \sum_f\g j_f\Theta_f
\ee
which is the Lorentzian Regge action for discrete GR if $\g j_f$ is identified to be the area. Such a result including the geometrical interpretation has the following meaning: if we consider perturbation expansion of the partial amplitude $A_j(\ck)$ developed from a vacuum, and let the vacuum be a spinfoam critical configuration $(j_f,g_{ve},z_{vf})$ equivalent to a globally oriented, time-oriented, nondegenerate geometry, the leading order of the perturbative expansion is an exponential of Regge action in GR. All the corrections to the Regge action from non-critical configurations in $A_j(\ck)$ is classified in the $1/\l$-corrections.

We refer to \cite{HZ} for the critical action at a generic spinfoam critical configuration corresponding to non-globally oriented, or time-nonoriented spacetime, as well as the degenerate situation. In all the unphysical situations, the critical action doesn't precisely give a Lorentzian Regge action, but the following modifications:
\begin{description}

\item[\underline{Time-Nonoriented}:] If the critical data doesn't correspond to a globally time-oriented simplicial spacetime, the critical action contains a term in addition to the Regge action 
\be
S_{\text{Time-Nonoriented }}=-i\eps\pi\sum_fj_fn_f
\ee
where $n_f=1$ if the time-orientation condition is violated at a face $f$, i.e. $\prod_{e\subset \partial f}\mu_e=-1$, otherwise $n_f=0$.

\item[\underline{Spacetime-Nonoriented}:] In general for a region with uniform spacetime orientation $\mathrm{sgn}(V_4)$ is either $+1$ or $-1$, the critical action is given by $\mathrm{sgn}(V_4)S$. Thus for a critical data with non-uniform $\mathrm{sgn}(V_4)$, one has to make a partition of the simplicial complex into regions $\{\calr_I\}_I$, such that $\mathrm{sgn}(V_4)$ is a constant on each region $\calr_I$. Then the critical action is given by a sum $\sum_{I}\mathrm{sgn}(V_4)_{\calr_I}S$   

\item[\underline{Euclidean}:] If the nondegenerate assumption Eq.\eqref{proddet} is violated by the critical data, and the critical data is classified to be Type-A, the critical action is an analog of Euclidean Regge action (with the modifications) $S=-i \mathrm{sgn}(V_4) \eps\lt[\sum_f j_f\Theta^E_f+\pi\sum_fj_fn_f\rt]$ if $j_f$ is identified to be the area. 

\item[\underline{Vector Geometry}:] If the nondegenerate assumption Eq.\eqref{proddet} is violated by the critical data, and the critical data is classified to be Type-B, the critical action can be written as
$S=-i\sum_fj_f\Phi_f$ where $\Phi_f$ is an 3-d rotation angle determined by the vector geometry.

\end{description}

\section{Conclusion}

In the paper, a new path integral representation of the Lorentzian EPRL spinfoam amplitude is derived from the group-representation-theoretic definition of the model in \cite{Carlo}. The new path integral representation, as well as the new spinfoam action, are the starting point to study the large spin asymptotics of the spinfoam amplitude via the stationary phase approximation.

Furthermore we develop the analysis of the spinfoam large spin asymptotics from \cite{HZ}. We clarity that in the large spin regime, there is an equivalence between the spinfoam critical configuration (with certain nondegeneracy assumption) and the classical Lorentzian simplicial geometry. Such an equivalence relation allows us to classify and count the spinfoam critical configurations by their geometrical interpretations. There are two types of solution-generating maps completely generating all the geometric critical configurations, where one type of the solution-generating maps is the (local) parity inversion firstly discussed in \cite{semiclassical}. 

On the other hand, the equivalence between spinfoam critical configuration and simplical geometry also allows us to define the notion of globally oriented and time-oriented spinfoam critical configuration. It is shown that only at a globally oriented and time-oriented spinfoam critical configuration, the leading order contribution of spinfoam large spin asymptotics gives precisely an exponential of Lorentzian Regge action. At all other critical configurations, spinfoam large spin asymptotics modifies the Regge action at the leading order approximation.

It is interesting to compare the spinfoam action proposed in the present paper and the one used in \cite{HZ}. Such an comparison is done in Appendix \ref{redundant}, where we argue the spinfoam action Eq.\eqref{SSS} proposed here is preferred since the other one in \cite{HZ} may in danger of some degeneracy. Such a degeneracy doesn't appear in the spinfoam action here. Through the comparison, we see the efficiency of the path integral representation defined here by a reduced number of degrees of freedom. The path integral representation proposed here may also be viewed as resulting from the one in \cite{HZ} by integrating out the $\xi_{ef}$ spinors (relating to the normal of a triangle). The variables in the path integral representation has the geometric meaning as the case in \cite{HZ}, which is clarified in the asymptotic analysis in this paper. 

Finally we emphasize that the present paper only deals with the spinfoam partial amplitude with a fixed spin configuration. The semiclassical analysis taking into account summing over spins is carried out in the companion papers \cite{statesum,han}, which leads to a 2-parameter expansion of the spinfoam amplitude.

\section*{Acknowledgements}

M.H and T.K would like to thank H. Haggard, S. Speziale, A. Riello, C. Rovelli, and M. Zhang for many helpful discussions. M.H also would like to thank Y. Ma for the invitation to visit the Center for Relativity and Gravitation, Beijing Normal University, where a part of this research work is carried out. The research leading to these results has received funding from the People Programme (Marie Curie Actions) of the European Union's Seventh Framework Programme (FP7/2007-2013) under REA grant agreement No. 298786.

\appendix

 \section{Algebraic Preliminaries: Principal Series of $\Slc$}\label{Slc}
 
 Following the book by W. R\"uhl \cite{ruhl}, we give here an overview of the principal series representations of  $\Slc$, realized on functions on $\Su$.
 
We consider the non-compact group $\Slc$ of complex matrices of determinant 1
 \begin{equation}
 SL(2,{\Bbb C})=\left\{g=\begin{pmatrix}a&c\cr b&d\end{pmatrix}\ a,b,c,d\in{\Bbb C}\quad ad-bc=1\right\}.
 \end{equation}
The principal series are infinite dimensional unitary irreducible representations of $\Slc$ indexed by a half integer  $k$ and a real parameter $\nu$. These representations are not equivalent except ${\cal H}_{k,\nu}\simeq {\cal H}_{-k,-\nu}$, so that we restrict our attention to non negative $k$.

${\cal H}_{k,\nu}$ is conveniently described in terms of functions on the compact group $\Su$,
 \begin{equation}
 SU(2)=\left\{h=\begin{pmatrix}\bar{v}&-\bar{u}\cr u&v\end{pmatrix}\ u,v\in{\Bbb C}\quad |u|^{2}+|v|^{2}=1\right\}.
\end{equation}
Complex valued functions on $\Su$ equipped with the $L^{2}$ scalar product derived from the Haar measure form a Hilbert space $L^{2}(\Su)$. Using the Peter-Weyl theorem, any such function can be expanded on the Wigner matrices
\begin{equation}
F(h)=\sum_{j\in{\Bbb N}/2,\atop m,n\in\left\{-j,-j+1,\dots,j\right\}} f_{m,n}^{j}D_{m,n}^{j}(h).\label{PeterWeyl}
\end{equation} 
The matrix elements obey the Schur orhogonality relations
\begin{equation}
\int_{SU(2)} dh\, \overline{D_{m,n}^{j}(h)}D_{m',n'}^{j'}(h)=\frac{1}{d_{j}}\delta^{j,j'}\delta_{m,n}\delta_{m',n'},\label{Schur}
\end{equation}
with $d_{j}=2j+1$. As a Hilbert space, ${\cal H}_{k,\nu}$  is the subspace of $L^{2}(\Su)$ made of functions obeying the covariance condition
\begin{equation}
F(\gamma h)=\mathrm{e}^{2\mathrm{i}k\phi}F(h)\quad\mbox{for any }\gamma=
\begin{pmatrix}\mathrm{e}^{\mathrm{i}\phi}&0\cr0&\mathrm{e}^{-\mathrm{i}\phi}\end{pmatrix}\in SU(2).\label{covariance}
\end{equation}
Since $D_{m,n}^{j}(\gamma)=\delta_{m.n}\mathrm{e}^{2\mathrm{i}m\phi}$, the covariance condition \eqref{covariance} restricts the summation in \eqref{PeterWeyl} to $m=k$, hence $j\geq k$ with $2j$ and $2k$ having the same parity. An orthonormal basis of ${\cal H}_{k,\nu}$ is therefore made of a subset of Wigner matrices, considered as functions of $h$,
\begin{equation}
\langle h|j,m\rangle_{{\cal H}_{k,\nu}}=\sqrt{d_{j}}\,D_{k,m}^{j}(h),\ \ \ m\in\{-j,\cdots,j\}, \ k\leq j,\label{basis}
\end{equation}
where we used the subscript ${\cal H}_{k, \nu}$ to distinguish this basis from the standard basis $|j,m\rangle$ of the spin $j$ representation of $\Su$. Consequently, as a representation of $\Su$ (an $\Su$ element acting by right multiplication on the argument of $F$), it decomposes as
\begin{equation}
{\cal H}_{k,\nu}=\mathop{\bigoplus}\limits_{j-k\,\in\,{\Bbb N}}{\cal H}_{j}\label{decompositionSU(2)}
\end{equation}
with ${\cal H}_{j}$ the standard spin $j$ representation of $\Su$. This is the action of the subgroup $\Su\subset \Slc$.  Note that $\Su$ does not see the real parameter $\nu$. 

To define the action of $\Slc$, let us first note that any $g\in \Slc$ can be written as
\begin{equation}
g=kh\label{decomposition}
\end{equation}
with $h\in SU(2)$ and $k\in K$, where $K$ is the $SL(2,{\Bbb C})$ subgroup
\begin{equation}
K=\left\{k=\begin{pmatrix}\lambda ^{-1}&\mu\cr0&\lambda\end{pmatrix}, \lambda\in{\Bbb C}^{\ast},\mu\in{\Bbb C}\right\}.
\end{equation}
Explicitly, this decomposition reads
\begin{equation}
\begin{pmatrix}a&c\cr b&d\end{pmatrix}=\begin{pmatrix}\lambda^{-1}&\mu\cr0&\lambda\end{pmatrix}\begin{pmatrix}\bar{v} &-\bar{u}\cr u&v\end{pmatrix}
\end{equation}
with
\begin{align}
\lambda&=\mathrm{e}^{\mathrm{i}\phi}(|b|^{2}+|d|^{2})^{1/2},&\mu&=a\overline{u}+c\overline{v},\cr
u&=\frac{\mathrm{e}^{-\mathrm{i}\phi}b}{(|b|^{2}+|d|^{2})^{1/2}},& v&=\frac{\mathrm{e}^{-\mathrm{i}\phi}d}{(|b|^{2}+|d|^{2})^{1/2}}.
\end{align}
and $\mathrm{e}^{\mathrm{i}\phi}$ an arbitrary phase. As an aside, let us note that the decomposition \eqref{decomposition} yields the isomorphism of cosets 
\begin{equation}
\Slc/K\simeq \Su/\mathrm{U}(1)\simeq S^{2}.
\end{equation}
Moreover, since $h=k^{-1}g$ is unitary, $(h^{-1})^{\dagger}=h$, it is useful to notice that
\begin{equation}
k^{-1}g=k^{\dagger}(g^{-1})^{\dagger}\label{kdagger}.
\end{equation}
To define the $\Slc$ action on ${\cal H}_{k,\nu}$, let us write for any $h\in\Su$ and $g\in \Slc$
\begin{equation}
hg=k_{g}(h)h_{g}(h)
\end{equation}
using the decomposition \eqref{decomposition} for $hg$, with $k_{g}(h)\in K$ and $h_{g}(h)\in \Su$. Then, the action of $g$ on $F\in{\cal H}_{k,\nu}$ reads\footnote{$k$ and $\nu$ are multiplied by 2 with respect to convention of \cite{ruhl}, for notational convenience}
\begin{equation}
g\cdot F[h]=\big[\lambda_{g}(h)\big]^{-k+\mathrm{i}\nu-1}\big[\overline{\lambda_{g}(h)}\big]^{k+\mathrm{i}\nu-1}F[h_{g}(h)]
\end{equation}
Thanks to the covariance condition \eqref{covariance}, the phase ambiguity in the decomposition \eqref{decomposition} is irrelevant. Therefore, we choose $\phi=0$ so that $\lambda_{g}(h)$ is real. In this case $\lambda_{g}^{2}(h)$ is nothing but the upper left corner of the matrix  $(k_{g}(h)^{-1})^{\dagger}k_{g}^{-1}(h)=h(g^{-1})^{\dagger}g^{-1}h^{\dagger}$,  so that
\begin{equation}
\lambda_{g}(h)=\Big[\langle\textstyle{\frac{1}{2},\frac{1}{2}}|h(g^{-1})^{\dagger}g^{-1}h^{\dagger}|\textstyle{\frac{1}{2},\frac{1}{2}}\rangle \Big]^{\frac{1}{2}}
\label{lambda}
\end{equation}
with $|j,m\rangle$ the standard basis of the spin $j$ representation of $SU(2)$. %In this case, $|\textstyle{\frac{1}{2},\frac{1}{2}}\rangle=\begin{pmatrix}1\cr 0\end{pmatrix}$.

\section{Critical Equations}\label{CE}

In this section we derive the critical equations $S'=0$ and $\Re(S)=0$. First of all, in order to apply Theorem \ref{asymptotics}(A) to the spinfoam action $S$, we must show that the spinfoam action $S$ satisfies $\Re(S)\leq0$. It can be proven by using Cauchy-Schwarz inequality $|\lag\chi,\psi\rag|^2\leq \lag\chi,\chi\rag\lag\psi,\psi\rag$:
\be
\Re(S)=\sum_{(e,f)}\ln\frac{\lt|\lag Z_{vef},Z_{v'ef}\rag\rt|^{2}}{ \lag Z_{v'ef},Z_{v'ef}\rag \lag Z_{vef},Z_{vef}\rag}
\leq0
\ee
The critical equation $\Re(S)= 0$ is equivalent to the condition that $Z_{vef},Z_{v'ef}$ are proportional to each other with complex coefficient, i.e.
\be
\frac{g^\dagger_{ve}z_{vf}}{\lt|\lt|Z_{vef}\rt|\rt|}=e^{i\a^f_{vv'}}\frac{g^\dagger_{v'e}z_{v'f}}{\lt|\lt|Z_{v'ef}\rt|\rt|}.\label{glue1}
\ee

There are two types of equation of motion, i.e. $\delta_{z_{vf}}S=0$ and $\delta_{g_{ve}}S=0$. First of all we compute the variation of $S$ with respect to the $\mathbb{CP}^1$-spinor $z_{vf}$. Given a spinor $z^\a=(z_0,z_1)^t$, $z^\a$ and $(Jz)^\a=(-\bar{z}_1,\bar{z}_0)^t$ is a basis of the space $\bbc^2$ of 2-component spinors, and $\lag z,Jz\rag=0$ with respect to the Hermitian inner product. The variation of $z_{vf}$ can be written in general by
\be
\delta z_{vf}=\eps_{vf} Jz_{vf}+\o_{vf} z_{vf}
\ee 
where $\eps,\o$ are complex number. Because of the gauge transformation $z_{vf}\mapsto\l z_{vf},\l\in\C$, we can choose a partial gauge fixing that $\lag z_{vf},z_{vf}\rag=1$, which gives $\lag \delta z_{vf},z_{vf}\rag=-\lag z_{vf},\delta z_{vf}\rag$. Thus we obtain $\o_{vf}=i\eta_{vf}$ with a real number $\eta_{vf}$. Then we have ($v=s(e)=t(e')$)
\be
\delta_{z_{vf}}S&=&2j_f\frac{\bar{\eps}_{vf}\lag J z_{vf},g_{ve}Z_{t(e)ef}\rag}{\lag Z_{vef},Z_{t(e)ef}\rag}+2j_f\frac{{\eps}_{vf}\lag g_{ve'}Z_{s(e')e'f},Jz_{vf}\rag}{\lag Z_{s(e')e'f},Z_{ve'f}\rag}\nonumber\\
&&+(i\g-1)j_f\frac{\bar{\eps}_{vf}\lag Jz_{vf},g_{ve}Z_{vef}\rag}{\lag Z_{vef},Z_{vef}\rag}+(i\g-1)j_f\frac{{\eps}_{vf}\lag g_{ve}Z_{vef},Jz_{vf}\rag}{\lag Z_{vef},Z_{vef}\rag}\nonumber\\
&&-(i\g+1)j_f\frac{\bar{\eps}_{vf}\lag Jz_{vf},g_{ve'}Z_{ve'f}\rag}{\lag Z_{ve'f},Z_{ve'f}\rag}-(i\g+1)j_f\frac{{\eps}_{vf}\lag g_{ve'}Z_{ve'f},Jz_{vf}\rag}{\lag Z_{ve'f},Z_{ve'f}\rag}
\ee 
which can be simplified by using Eq.\eqref{glue1}
\be
\delta_{z_{vf}}S&=&(i\g+1)j_f\frac{\bar{\eps}_{vf}\lag Jz_{vf},g_{ve}Z_{vef}\rag}{\lag Z_{vef},Z_{vef}\rag}+(i\g-1)j_f\frac{{\eps}_{vf}\lag g_{ve}Z_{vef},Jz_{vf}\rag}{\lag Z_{vef},Z_{vef}\rag}\nonumber\\
&&-(i\g+1)j_f\frac{\bar{\eps}_{vf}\lag Jz_{vf},g_{ve'}Z_{ve'f}\rag}{\lag Z_{ve'f},Z_{ve'f}\rag}-(i\g-1)j_f\frac{{\eps}_{vf}\lag g_{ve'}Z_{ve'f},Jz_{vf}\rag}{\lag Z_{ve'f},Z_{ve'f}\rag}.
\ee
Then $\delta_{z_{vf}}S$ implies that
\be
\frac{\lag Jz_{vf},g_{ve}Z_{vef}\rag}{\lag Z_{vef},Z_{vef}\rag}=\frac{\lag Jz_{vf},g_{ve'}Z_{ve'f}\rag}{\lag Z_{ve'f},Z_{ve'f}\rag}
\ee
Because $z_{vf}$ and $Jz_{vf}$ is a basis of the space $\bbc^2$ of 2-component spinors, and ${\langle z_{vf},g_{ve}Z_{vef}\rangle}{|| Z_{vef}||^{-2}}={\langle z_{vf},g_{ve'}Z_{ve'f}\rangle}{|| Z_{ve'f}||^{-2}}$ is a trivial identity, we obtain the equation of motion
\be
\frac{g_{ve}g_{ve}^\dagger z_{vf}}{\lag Z_{vef},Z_{vef}\rag}=\frac{g_{ve'}g_{ve'}^\dagger z_{vf}}{\lag Z_{ve'f},Z_{ve'f}\rag}\label{glue2}.
\ee

We compute the equation of motion corresponding to the variation with respect to the group variables $g_{ve}$. Let's assume that $v=s(e)$, the case $v=t(e)$ follows similarly. Consider the following variation of the group variables:
\begin{equation}
g_{ve}\rightarrow g_{ve}-\delta g_{ve}\quad\mathrm{with}\quad \delta g_{ve}=g_{ve}A^\dagger_{ve},
\end{equation}
where $A_{ve}\in \slc$ is a traceless $2\times 2$ complex matrix. Then we compute the following variations:
\be
\delta\lt(\ln\lag g_{ve}^\dagger z_{vf}, g_{ve}^\dagger z_{vf}\rag\rt)=
-\frac{\lag g_{ve}^\dagger z_{vf}\big|A_{ve}+A_{ve}^{\dagger}\big| g_{ve}^\dagger z_{vf}\rag}{\lag g_{ve}^\dagger z_{vf}, g_{ve}^\dagger z_{vf}\rag}
\ee
and
\be
&&\delta\lt(\ln\lag g_{v'e}^\dagger z_{v'f}, g_{ve}^\dagger z_{vf}\rag\rt)=
-\frac{\lag g_{v'e}^\dagger z_{v'f}\big|A_{ve} \big|g_{ve}^\dagger z_{vf}\rag}{\lag g_{v'e}^\dagger z_{v'f}, g_{ve}^\dagger z_{vf}\rag}
=-\frac{\lag g_{ve}^\dagger z_{vf}\big|A_{ve} \big|g_{ve}^\dagger z_{vf}\rag}{\lag g_{ve}^\dagger z_{vf}, g_{ve}^\dagger z_{vf}\rag}\nonumber\\
&&\delta\lt(\ln\lag g_{ve}^\dagger z_{vf}, g_{v'e}^\dagger z_{v'f}\rag\rt)=
-\frac{\lag g_{ve}^\dagger z_{vf}\big|A_{ve}^\dagger \big|g_{v'e}^\dagger z_{v'f}\rag}{\lag g_{ve}^\dagger z_{vf}, g_{v'e}^\dagger z_{v'f}\rag}
=-\frac{\lag g_{ve}^\dagger z_{vf}\big|A_{ve}^\dagger \big|g_{ve}^\dagger z_{vf}\rag}{\lag g_{ve}^\dagger z_{vf}, g_{ve}^\dagger z_{vf}\rag}
\ee
once we have used Eq.\eqref{glue1} to replace $g_{v'e}^\dagger z_{v'f}$ in terms of $g_{ve}^\dagger z_{vf}$ .

We introduce the incidence matrix $\eps_{ef}(v)$ such that
\begin{equation}
\eps_{ef}(v)=
\begin{cases}
\:\:\:0&\text{if $v\notin\partial f$}\cr
\:\:\:1&\text{if $v=t(e)$ with $e\in\partial f$}\cr
-1&\text{if $v=s(e)$ with $e\in\partial f$}
\end{cases}
\end{equation}
$\eps_{ef}(v)$ satisfies the following relations:
\be
\eps_{ef}(v)=-\eps_{e'f}(v)\ \ \ \ \text{and}\ \ \ \ \eps_{ef}(v)=-\eps_{ef}(v').
\ee

Then, the variation of the action reads
\be
\delta_{g_{ve}}{S}=-\sum_{f}j_{f}\eps_{ef}(v)\frac{ \lag Z_{vef}\lt[A_{ve}-A_{ve}^{\dagger}+\mathrm{i}\gamma(A_{ve}+A_{ve}^{\dagger})\rt]Z_{vef}\rag
}{\lag Z_{vef},Z_{vef}\rag}
\ee
Since $A_{ve}=\sum_{j=1}^3\theta_{ve}^j\sig_j$ where $\theta_{ve}^j$ are complex parameters and $\vec{\sig}$ are Pauli matrices. Then the action $S$ is stationary with respect to the variation of $g_{ve}$ if and only if
\be
\sum_{f}j_{f}\eps_{ef}(v)\frac{ \lag Z_{vef}\ \vec{\sig}\ Z_{vef}\rag
}{\lag Z_{vef},Z_{vef}\rag}=0. \label{closure}
\ee
Finally note that if we consider $v=t(e)$ instead of $v=s(e)$, we obtain the same equation thanks to Eq.\eqref{glue1}.

Given a normalized 2-spinor $z$ (recall our partial gauge fixing condition $\lag z,z\rag=1$), it determines a null vector $\iota(z)\equiv\frac{1}{\sqrt{2}}(1,\hat{n}_z)$ via $zz^\dagger=\frac{1+\hat{n}_z\cdot\vec{\sig}}{2}$ where $\hat{n}_z\in S^2$ is a unit 3-vector. The $\Slc$ action $g zz^\dagger g^\dagger,g\in\Slc$ gives the Lorentz transformation of the null vector $\iota(z)$. If we make the following normalzation
\be
\frac{g zz^\dagger g^\dagger}{\lag g z,g z\rag}=\frac{1+g(\hat{n}_z)\cdot\vec{\sig}}{2}\ \ \ \text{where}\ \ \ 
g(\hat{n}_z)=\frac{\lag g z\ \vec{\sig}\ g z\rag}{\lag g z,g z\rag}
\ee
$g(\hat{n}_z)$ gives the action of $\Slc$ on 2-sphere. Therefore the equation of motion Eq.\eqref{closure} implies
\be
g^\dagger_{ve}\lt(\sum_f\eps_{ef}(v)j_f\hat{n}_{vf}\rt)=0\ \ \ \text{i.e.}\ \ \ \sum_f\eps_{ef}(v)j_f\hat{n}_{vf}=0
\ee
where $\hat{n}_{vf}\equiv\hat{n}_{z_{vf}}$. We call this relation the closure condition. The closure condition determines uniquely a convex geometrical tetrahedron whose face triangles $f$ have the oriented areas $j_f\hat{n}_{vf}$, up to a SO(3) rotation and a uniformly rescaling \cite{shape,polyhedra}.

\section{Proof of Lemma \ref{factorization}}\label{5.1}

The result in Lemma \ref{factorization} was firstly shown in \cite{CF}. Let's construct the following relation:
\be
0 =\b_{km}(v)\sum_{j=1}^5\b_{lj}(v)N_{e_j}(v)-\b_{lm}(v)\sum_{j=1}^5\b_{kj}(v)N_{e_j}(v)=\sum_{j\neq m}\Big[\b_{km}(v)\b_{lj}(v)-\b_{lm}(v)\b_{kj}(v)\Big]N_{e_j}(v)
\ee
Since we assume the nondegeneray condition Eq.\eqref{proddet}, any four of the five $N_{e}(v)$ are linearly independent. Thus
\be
\b_{km}(v)\b_{lj}(v)=\b_{lm}(v)\b_{kj}(v)
\ee
Let us pick one $j_0$ for each 4-simplex, and choose $l=j=j_0$. We obtain the following because $\b_{jj}(v)$ is nonzero
\be
\b_{km}(v)=\frac{\b_{kj_0}(v)\b_{mj_0}(v)}{\b_{j_0j_0}(v)}.
\ee
Therefore we have the factorization of $\b_{ij}(v)$
\be
\b_{ij}(v)=\varsigma(v)\b_i(v)\b_j(v)
\ee
where $\b_j(v)={\b_{jj_0}(v)}\big/\sqrt{\lt|\b_{j_0j_0}(v)\rt|}$ and $\varsigma(v)=\mathrm{sgn}(\b_{j_0j_0}(v))$ which is a constant sign within a 4-simplex $\sig_v$. So far it seems that the value of $\b_j(v)$ and ${\varsigma}(v)$ depends on the choice of the dual edge $e_{j_0}$ connecting to $v$. However we show in the following analysis that such a dependence doesn't affect the geometrical interpretation.

\section{Degenerate Geometrical Interpretation}\label{DEG}

The previous discussion respects the nondegeneracy assumption Eq.\eqref{proddet}. In this section we briefly discuss the case when Eq.\eqref{proddet} is violated, i.e. we consider a \emph{lower dimensional} subspace of spinfoam configurations $(j_f,g_{ve},z_{vf})$ given by 
\be
\det\Big(N_{e_1}(v),N_{e_2}(v),N_{e_3}(v),N_{e_4}(v)\Big)=0\label{deg}
\ee
for some choices of 4 different edges. We can construct the bivectors by using the group variables satisfying Eq.\eqref{deg} as we did before. However it is shown in \cite{semiclassical} that the bivectors in each simplex are contained in a 3-dimensional subspace orthogonal to $(1,0,0,0)^t$, once Eq.\eqref{deg} is satisfied. It means that all the unit vectors $N_e(v)$ are parallel and equal to $(1,0,0,0)^t$ and the critical group variables $g_{ve}$ belong to SU(2). Thus the critical equation Eq.\eqref{2} becomes a trivial equation, while the other critical equations Eqs.\eqref{1} and \eqref{3} reduce to
\be
g_{ve}^{-1}z_{vf}=e^{i\a_{vv'}^f}g_{v'e}^{-1}z_{v'f},\ \ \text{and}\ \ \ \sum_{f}j_{f}\eps_{ef}(v){ \big\langle  g_{ve}^{-1}z_{vf}\big| \vec{\sig}\big| g_{ve}^{-1}z_{vf}\big\rangle }=0,
\ee
if we partially gauge fix $\lag z_{vf},z_{vf}\rag=1$. Accordingly in terms of the auxiliary spinor $\zeta_{ef}$, the critical equations Eq.\eqref{zeta1}, \eqref{zeta2}, and \eqref{zeta3} reduce to
\be
{g_{ve}\zeta_{ef}}=e^{i\phi_{eve'}}g_{ve'}\zeta_{e'f},\ \ \text{and}\ \ \ 
\sum_{f}\eps_{ef}(v)j_{f}{ \lag \zeta_{ef}\ \vec{\sig}\ \zeta_{ef}\rag}\equiv \sum_{f}\eps_{ef}(v)j_{f}\hat{n}_{ef}
=0\label{eucl}
\ee
Which are essentially the critical equations of large spin asymptotics in Euclidean EPRL spinfoam amplitude (studied in \cite{semiclassical,HZ}) or in SU(2) BF theory (studied in \cite{asym15j}). The result of analysis can be summarized in the following: There are two types of solutions of the critical equations when the degenercy Eq.\eqref{deg} is satisfied. 

\begin{description}

\item[\underline{Type-A}:] Given a solution $(j_f,g_{ve},\zeta_{ef})$ of Eq.\eqref{eucl}, it may determine uniquely another solution $(j_f,g'_{ve},\zeta_{ef})$ in each individual 4-simplex with $g'_{ve}\neq g'_{ve}$ \cite{semiclassical}. Because of the uniqueness, we can write the critical solution to be $(j_f,(g_{ve},g'_{ve}),\zeta_{ef})$ with $g_{ve},g'_{ve}\in\mathrm{SU(2)}$. Then it turns out that the critical data $(j_f,(g_{ve},g'_{ve}),\zeta_{ef})$ is equivallent to a Euclidean (discrete) geometrical data:
\be
(j_f,(g_{ve},g'_{ve}),\zeta_{ef})\Longleftrightarrow (\pm_v E^E_\ell(v),\eps,\eps_e(v)).
\ee
$E^E_\ell(v)$ is a nondegenerate discrete cotetrad of Euclidean geometry, where the nondegeneracy is implied by $g_{ve}\neq g'_{ve}$. Here $\eps_e(v)=\mathrm{sgn}\lt(U^E_e(v)N^E_e(v)\rt)$ ($N^E_e(v)=(g_{ve},g'_{ve})\act (1,0,0,0)^t$) has to appear in order to establish the equivalence since the notion of future/past is not an invariant under the action of SO(4), as we mentioned in a previous footnote.

Note that such a type of critical configuration can appear from any Regge-like spin configuration, which implies the nondegeneracy of the Lorentzian geometry, since all the edge-vectors are space-like, as a property shared by both the Lorenzian and Euclidean 4-simplices. Such a type of critical configuration can also appear from some Non-Regge-like spin configurations in the Lorenzian sense, which however is Regge-like in the Euclidean sense, i.e. there exists edge-lengths $s_\ell^E$ such that $\g j_f$ are the Euclidean areas given by the edge-lengths\footnote{The (Lorentzian) Regge-like spin configurations considered previously is a proper subset of Euclidean Regge-like spin configurations.}.

\item[\underline{Type-B}:] There exist the critical solutions $(j_f,g_{ve},\zeta_{ef})$ which don't give any other solution, or namely, give uniquely another solution $(j_f,g'_{ve},\zeta_{ef})$ with $g'_{ve}=g_{ve}$. In such a case, the geometrical 4-simplex is degenerate even in the Euclidean sense. The corresponding geometry is called a vector geometry \cite{semiclassical}. 

\end{description}

%Here we propose a simple way to remove from the spinfoam amplitude the critical configurations discussed in the last section, which violates the nondegeneracy assumption Eq.\eqref{proddet}. There are existing method to remove these critical configurations by imposing a projective operator in the spinfoam vertex amplitude \cite{Engle}. However the strategy we propose here is to implement a local measure in the path integral representation of spinfoam model.

\section{Comparison of Spinfoam Actions}\label{redundant}

In this section, we compare the path integral formulation of EPRL spinfoam amplitude proposed here and the one proposed in the first reference of \cite{HZ}, and we give an argument why the path integral formulation proposed here is preferred. 

The path integral formulation in \cite{HZ} is constructed by consistently gluing the vertex amplitude in \cite{semiclassical}. The spinfoam amplitude can be written as
\be
A(\ck)=\sum_{j_{f}}\prod_{f}\mu \left( j_{f}\right) \prod_{\left( v,e\right)}\int_{\Slc}
\rmd g_{ve} \prod_{\left( e,f\right)}\int_{S^2} \rmd \xi_{ef}\prod_{v\in \partial f}\int_{\mathbb{CP}^{1}}\left( \frac{\dim(j_f)}{\pi }\Omega
_{vf}\right) e^{S'\lt[j_f,g_{ve},z_{vf},\xi_{ef}\rt]}
\ee%
where the spinfoam action $S'$ in \cite{HZ} is given by $S'=\sum_f S'_f$ and 
\begin{equation}
S'_{f}=\sum_{v\in f}S'_{vf}=\sum_{v\in f}\left( j_{f}\ln \frac{\left\langle
\xi_{ef},Z_{vef}\right\rangle ^{2}\left\langle Z_{ve^{\prime }f},\xi_{e^{\prime
}f}\right\rangle ^{2}}{\left\langle Z_{vef},Z_{vef}\right\rangle
\left\langle Z_{ve^{\prime }f},Z_{ve^{\prime }f}\right\rangle }+i\gamma
j_{f} \ln \frac{\left\langle Z_{ve^{\prime }f},Z_{ve^{\prime
}f}\right\rangle }{\left\langle Z_{vef},Z_{vef}\right\rangle }\right).
\end{equation}
The spinfoam action $S$ proposed here may be viewed as an ``effective action'' from the above path integral by integrating out the $\xi_{ef}$-variables ($\xi_{ef}$ is a normalized 2-spinor). 

If we perform the (generalized) stationary phase analysis to the action $S'$, we obtain the following equations equivalent to $\Re(S')=0$:
\begin{equation}
\xi_{ef}=\frac{e^{i\phi^f _{ev}}}{\left\Vert Z_{vef}\right\Vert }Z_{vef}.\label{ReS'}
\end{equation}%
with the undetermined phase $e^{i\phi^f _{ev}}$. However these equations are actually equivalent to the equation of motion $\delta_{\xi_{ef}}S=0$. Indeed if we perform the variation of $S'$ with respect to normalized spinor $\xi_{ef}$ by using $\delta\xi_{ef}^\a=\o_{ef}( J\xi_{ef})^\a+i\eta_{ef}\xi^\a_{ef}$ for complex infinitesimal parameter $\o\in\bbc$ and $\eta\in\mathbb{R}$, the variation of the action gives 
\begin{eqnarray}
\delta _{\xi_{ef}}S =j_{f}\left( 2\frac{\bar{\o}_{ef}\left\langle
J\xi_{ef},Z_{vef}\right\rangle }{\left\langle \xi_{ef},Z_{vef}\right\rangle }+2%
\frac{\o_{ef} \left\langle Z_{v^{\prime }ef},J\xi_{ef}\right\rangle }{%
\left\langle Z_{v^{\prime }ef},\xi_{ef}\right\rangle }\right) 
\end{eqnarray}%
Since $\xi_{ef}$ and $J\xi_{ef}$ is an orthonormal basis in $\C^2$ with the Hermitian inner product, $\delta _{\xi_{ef}}S=0$ is equivalent to the condition that $\xi_{ef}\propto_{\C}Z_{vef}$, which is nothing but Eq.\eqref{ReS'}. Therefore we have the following equivalence relation
\be
\Re(S')=0\Longleftrightarrow\delta _{\xi_{ef}}S'=0\Longleftrightarrow\xi_{ef}\propto_{\C}Z_{vef}
\ee
for the action $S'$.

When we consider the variation of $S'$ with respect to the group variable $g_{ve}$ and insert in the equation of motion $\xi_{ef}\propto_{\C}Z_{vef}$, we find the variations by the rotation part of $g_{ve}$ and the variations by the boost part give exactly the same equation, i.e.
\be
\delta_{g_{ve}}^RS'=0\Longleftrightarrow\delta_{g_{ve}}^BS'=0\Longleftrightarrow\sum_fj_f\lag\xi_ef|\vec{\sig}|\xi_{ef}\rag=0\ \ \text{when}\ \ \delta _{\xi_{ef}}S'=0\label{RB}
\ee
In this way the 6 equations of motion $\delta_{g_{ve}}S'=0$ reduce to 3 equations (closure condition). Equivalently, if we make the variation along the direction defined by $\vec{K}+\g \vec{J}\in\slc$ ($\vec{K},\vec{J}$ are boost and rotation generators), we obtain trivial equations when $\xi_{ef}\propto_{\C}Z_{vef}$ is satisfied, i.e.
\be
\delta_{g_{ve}}^{\vec{K}+\g \vec{J}}S'=0\Longleftrightarrow0=0\ \ \ \text{when}\ \ \ \delta _{\xi_{ef}}S'=0.
\ee 

A linearization of the equation of motion $\delta_xS'[j,x]=0$, $x=(g_{ve},z_{vf},\xi_{ef})$ gives
\be
H_{x,x'}\cdot\delta x'=-\delta_j\delta_xS'[j,x]\cdot\delta j
\ee
where $H_{x,x'}=\delta_x\delta_{x'}S'[j,x]\big|_{x_0}$ is the Heissian matrix evaluated at a given critical solution. Then the linearized version of Eq.\eqref{RB} implies that
\be
H_{g^R, x}\cdot\delta x=-\delta_j\delta_{g^R}S'\cdot\delta j\ \ \text{and}\ \ H_{g^B, x}\cdot\delta x=-\delta_j\delta_{g^B}S'\cdot\delta j
\ee
is the same equation when we insert in the linearized version of $\delta _{\xi_{ef}}S'=0$ 
\be
H_{\xi, x}\cdot\delta x=-\delta_j\delta_{\xi}S'\cdot\delta j.
\ee
From this fact, the Hessian matrix for the action $S'$ may be in danger of degenercy, while such a degeneracy may not correspond to any gauge freedom of the action $S'$.

For each edge $e$ there are 12 degrees of freedom from $g_{ve}$ and $g_{v'e}$. The degeneracy shown here reduces by half the degrees of freedom from each of $g_{ve}$ and $g_{v'e}$. The SU(2) gauge invariance $g_{ve}\mapsto g_{ve}h^\dagger_e,\xi_{ef}\mapsto h_e\xi_{ef}$ of $S'$ reduces another 3 degrees of freedom. There are only 3 degrees of freedom left, which gives the closure condition $\sum_fj_f\hat{n}_{ef}=0$ as 3 (real) equations.

The path integral formulation with the action $S$ proposed here doesn't suffer this problem. Although we still have
\be
\delta_{g_{ve}}^RS'=0\Longleftrightarrow\delta_{g_{ve}}^BS'=0\Longleftrightarrow Eq.\eqref{3}
\ee
as 3 real equations corresponding to closure condition, the derivation of such an equivalence uses $\Re(S)=0$ instead of any equation of motion. The linearization of $\Re(S)=0$ doesn't relate to the Hessian matrix. Thus one cannot conclude a degeneracy of Hessian matrix\footnote{Rigorously speaking, it is a conjecture that the Hessian of $S$ proposed here is nondegenerate at the critical point discussed in this paper, especially at the critical points of nondegenerate geometry. However it is a reasonable conjecture for $S$ since there is no degeneracy of Hessian observed by comparing the number of equations of motion to the number of degree of freedom. But surely such an observation doesn't in general exclude all the degeneracy of Hessian, and moreover, we don't expect the Hessian is nondegenerate at all critical points. Even in the asumptotics of a single vertex amplitude, the Heissian is degenerate if any tetrahedron is degenerate (see the explicit expression of Hessian in \cite{semiclassical,HZ}). The critical data with degenerate tetrahedra is not discussed in the present paper. }. Therefore we see that the difficulty of the action $S'$ in \cite{HZ} precisely comes from the fact that $\Re(S')=0$ is equivalent to the equation of motion $\delta_{\xi_{ef}}S'=0$. The difficulty is resolved by integrating out $\xi_{ef}$-variables, which is achieved at the beginning of this paper for deriving the spinfoam action $S$.

\end{document}